\documentclass[useAMS,amssymb,epsfig,usegraphicx,twocolumn,usenatbib]{mn2e}
\def\aj{AJ}
\def\araa{ARA\&A}
\def\apj{ApJ}
\def\apjl{ApJ}
\def\aap{A\&A}
\def\mnras{MNRAS}
\def\nat{Nature}
\def\na{NewA}

\usepackage[usenames,dvips]{color}

\newif\ifAMStwofonts

\makeatother

\title[Spin-up of low mass classical bulges]
{Spin-up of low mass classical bulges in barred galaxies}

\author[Saha et al.]
{Kanak Saha\thanks{E-mail:saha@mpe.mpg.de}, Inma Martinez-Valpuesta \& Ortwin Gerhard\\
Max-Planck-Institut f\"ur Extraterrestrische Physik, Giessenbachstraße, D-85748 Garching, Germany}
 
\begin{document}

\date{Accepted xxxx Month xx. Received xxxx Month xx; in original form
2011 Nov. 24} 
\pagerange{\pageref{firstpage}--\pageref{lastpage}} \pubyear{2009}
\maketitle

\label{firstpage}

\begin{abstract}
Secular evolution is one of the key routes through which galaxies evolve along the 
Hubble sequence. Not only the disk undergoes morphological and kinematic changes, 
but also a preexisting classical bulge may be dynamically changed by the secular 
processes driven primarily by the bar. We study the influence of a growing bar on 
the dynamical evolution of a low mass classical bulge such as might be present in 
galaxies like the Milky Way. Using self-consistent high resolution {\it N}-body 
simulations, we study how an initially isotropic non-rotating small classical bulge 
absorbs angular momentum emitted by the bar. The basic mechanism of this angular 
momentum exchange is through resonances and a considerable fraction of the angular 
momentum is channeled through Lagrange point (-1:1) and ILR (2:1) orbits. In the phase 
of rapid dynamical growth, also retrograde non-resonant orbits absorb significant 
angular momentum. As a result of this angular momentum gain, the initially non-rotating 
classical bulge transforms into a fast rotating, radially anisotropic and triaxial object, 
embedded in the similarly fast rotating boxy bulge formed from the disk. Towards the end 
of the evolution, the classical bulge develops cylindrical rotation. By that time, its 
inner regions host a "classical bulge-bar" whose distinct kinematics could serve as 
direct observational evidence for the secular evolution in the galaxy. Some 
implications of these results are discussed briefly.

\end{abstract}


\section{Introduction}
\label{sec:introduc}
In the hierarchical structure formation scenario \citep{WhiteRees1978, FallEfstathiou1980} mergers 
have played a strong role in forming and shaping galaxies. One of the common product of major mergers 
are the classical bulges \citep{Kauffmanetal1993, Baughetal1996, Hopkinsetal2009} which are the 
central building blocks in spiral galaxies. There have been a couple of other mechanisms suggested
for the formation and growth of classical bulges e.g., monolithic collapse of primordial gas clouds 
\citep{Eggenetal1962}, the coalescence of giant clumps in gas-rich primordial galaxies 
\citep{Immelietal2004, Elmegreenetal2008}, 
multiple minor mergers \citep{Bournaudetal2007, Hopkinsetal2010}, accretion of small companions or 
satellites \citep{Aguerrietal2001}. The classical bulges formed via these processes seem to have little 
rotation as compared to the random motion. On the other hand, various observational measurements have 
confirmed that classical bulges in spiral galaxies possess rotation 
\citep{KI1982, Cappellarietal2007} about their minor axis and in most cases in the same sense as the 
disk rotates. It is also known that classical bulges rotate faster than elliptical galaxies and that often 
their rotation velocities are comparable to that of an isotropic oblate rotator model \citep{Binney1978}. 
So the origin of the systematic rotational motion observed in the classical bulges remains unclear. 

The photometric and kinematic properties of classical bulges as well as their origin are quite
distinct from those of the other class of bulges, the boxy/peanut and disk-like bulges. It is well 
known that the surface brightness profiles in classical bulges follow a sersic law $\mu(r) \sim r^{1/n}$ 
with sersic index $n\sim 4$. Whereas the sersic indices in boxy and disk-like bulges are, in general, 
low with $n \le 2$; so that their surface brightness profiles follow roughly an exponential distribution, 
see \cite{KormendyKennicut2004,Combes2009} for extensive reviews. The kinematics of bulges are well 
illustrated in the $v/\sigma - \epsilon$ plot \citep{KI1982, K1982} which clearly demonstrates the 
distinction in the kinematic properties of ellipticals, classical bulges and boxy bulges and brings out 
the fact that in terms of their rotational support, classical bulges fall in between ellipticals and 
boxy/disk-like bulges. Since the boxy as well as disk-like bulges are thought to have formed from disk 
material, the source of their angular momentum is known.   

The classical bulges formed early through major mergers and violent relaxation can subsequently accrete material quiescently
 as a result of which a disk grows inside-out \citep{MoMaowhite1998, Katzetal2003, SpringelHernquist2005, Keresetal2009}.  
The gas accretion may facilitate the disk to grow sufficiently for the disk self-gravity to dominate the 
internal dynamics. Eventually, a bar and/or spiral arms form in the disk and initiate 
secular processes in the galaxy. Indeed, bars are quite common in disk galaxies, about $2/3$ of the disk galaxies 
host a strong stellar bar in their central region \citep{Laurikainenetal2004, MarinovaJogee2007, MenendezDelmestreetal2007}. 
Therefore one might expect the disk to go bar unstable also in galaxies with preexisting classical bulges. In fact,
classical, boxy and disk-like bulges could co-exist \citep{Athanassoula2005, Erwin2008,Gadotti2009,Nowaketal2010} in 
a single galaxy, although the observational identification of the several components could be difficult. 
In the Milky Way, an upper limit on the mass of a classical bulge ($\sim 8 \%$ of the disk mass) has been set
by modelling the kinematics from the Bulge Radial Velocity Assay (BRAVA) data \citep{Shenetal2010}. But there is evidence for a 
metallicity gradient above the Galactic plane \citep{zoccalietal2008,zoccali2010} which is taken as an indication
 for the existence of a classical bulge in our Galaxy. It is therefore important to understand the 
dynamical interaction between preexisting classical bulges and bars in barred galaxies.    

In this paper, we investigate in considerable detail the interaction of a bar and a low mass classical bulge via 
a high resolution {\it N}-body simulation of a galaxy consisting of a live disk, bulge and dark matter halo, and follow the 
evolution of the dynamical structure and kinematics of the small classical bulge. We find that its dynamical evolution is 
strongly connected to the growth of the bar which forms spontaneously in the disk. 
During the secular evolution, the structure and kinematics of the bulge are altered significantly, developing an 
interesting and complex rotation structure; in particular, cylindrical rotation (which is considered as a typical proxy 
of boxy bulge) appears in the inner region of the classical bulge.   
 
The paper is organized as follows. Section~\ref{sec:barbulgeinteraction} summarizes the basics of bar-bulge interaction. 
Section~\ref{sec:galaxymodels} outlines the initial galaxy model and set up for the {\it N}-body simulation. The bar evolution 
and boxy bulge formation are described in Section~\ref{sec:barevolution}. Section~\ref{sec:Lzexchange} describes, in detail, 
the angular momentum exchange between the bar and the classical bulge. The evolution of the classical bulge, its structure
and kinematics are presented in Section~\ref{sec:bulgestructure}. Discussion and conclusions are contained in 
Sections~\ref{sec:discussion} and \ref{sec:concl} respectively. In the text, by bulge, we mean a classical bulge unless
mentioned otherwise.
 
\section{Bar-Bulge interaction}
\label{sec:barbulgeinteraction}
As we have seen (in Section~\ref{sec:introduc}), the possible co-existence of a bar and a classical bulge might be 
rather common in present day disk galaxies, and thus they are bound to interact gravitationally. In fact, a preexisting 
classical bulge in the disk has a strong influence on the formation and growth of the bar itself. For the swing amplification to 
work, one needs to keep alive the feedback loop through which a set of trailing waves traveling through the 
center are transformed into leading waves. This is possible as long as there is no inner 
Lindblad resonances (ILRs). A highly centrally concentrated bulge can shield the center by putting an 
ILR barrier and thus cutting the feedback loop which in turn could hinder the growth of the bar 
\citep{SellwoodEvans2001}. However, various non-linear processes are probably active in real galaxies 
which would destroy the ILR barrier and eventually lead to the formation of a bar 
\citep{Widrowetal2008, DubinskiBerentzen2009}. 

Once a bar is formed, it takes over the dynamics in the central region of the disk and starts interacting with the 
stellar bulge and dark matter halo through exchange of angular momentum. Based on the work of 
\cite{Lynden-BellKalnajs1972}, hereafter LBK72, it has been emphasized by several authors 
\citep{TremaineWeinberg1984, Weinberg1985, HernquistWeinberg1992, 
DebattistaSellwood2000, WeinbergKatz2002, Athanassoula2002, SellwoodDebattista2006,DubinskiBerentzen2009} that the resonant
interaction plays a significant role in the angular momentum transfer between the bar and the dark matter halo.
It has been suggested by \cite{HernquistWeinberg1992, Athanassoula2003, WeinbergKatz2007a} that the 
same underlying mechanism could as well apply between the bar and the spheroid and in particular, 
\cite{Athanamisi2002} have studied how the shape of a bulge would change in response to a growing bar. 

Although the dynamical interaction between a growing bar and a bulge and their subsequent 
evolution can be best studied via {\it N}-body simulations, an analytic understanding is 
required to complement this. Following LBK72, it can be shown that during the bar-bulge 
interaction the time rate of change of angular momentum of a classical bulge, whose distribution 
function ($F_b$) is descibed by a King model, is always positive and can be written as  

\begin{equation}
\dot{L}_{z,b} \sim \Omega_B \times |\psi_{lmn}|^2 \times F_b/\sigma_b^2 > 0,
\end{equation}
\noindent where $\psi_{lmn}$ and $\Omega_B$ are the Fourier amplitude and pattern speed 
of a non-responsive bar potential and $\sigma_b$ is the velocity dispersion of the bulge stars. 
So at a given resonance, the angular momentum gained by the bulge depends on the strength
of the bar and is inversely proportional to the square of bulge velocity dispersion, implying a hotter bulge
will absorb less angular momentum provided other conditions remain unchanged.
However, in real galaxies, the angular momentum transfer is more difficult to determine, because the time rate
of change of bar's angular momentum involves the change in its pattern speed, moment of inertia, and
 in the angular momentum associated with any internal circulation \citep{villavargasetal2009} 
within the bar. In Section~\ref{sec:Lzexchange}, we show the angular momentum transfer between the 
bar and the bulge in our simulation using orbital spectral analysis.

\section{galaxy model and {\it N}-body simulation}
\label{sec:galaxymodels}
An equilibrium model for a disk galaxy is constructed using the self-consistent bulge-disk-halo 
model of \citet{KD1995}. Their prescription provides nearly exact solutions of the collisionless 
Boltzmann and Poisson equations which are suitable for studying disk stability related problems. 
 All the components in our model are live (i.e., the gravitational potential of each 
component can respond to an external or internal perturbation) and, hence, provide a 
realistic representation for the structure and evolution of the galaxy. 
Below we briefly describe each component of the model. For more details, the reader 
is referred to \citet{KD1995}.

The disk distribution function is constructed using the approximate third integral 
given by $E_z = \frac{1}{2} v_z^2 +\Psi(R,z) - \Psi(R,0)$, the energy of the vertical 
oscillations. This third integral is approximately conserved for orbits near the disk 
mid-plane. The radial density of the disk is approximately exponential with a truncation, 
and the square of the radial velocity dispersion follows the same exponential radial 
decline with a scale length same as the disc scale length. The vertical structure of the 
disk is approximately isothermal, with the scale height set by the vertical velocity dispersion 
and vertical potential gradient. The volume density of the axisymmetric disk is given by

\begin{equation}
\rho_d(R,z) = \frac{M_d}{8\pi h_z R_d^2} e^{-R/R_d} \mathop{\mathrm{erfc}}\left(\frac{R - R_{out}}{\sqrt{2}(R_{out} -R_{trun})}\right) f_d(z),
\end{equation}

\noindent where $f_d(z) = \exp(-0.8676 \Psi_z(R,z)/\Psi_z(R,h_z))$ with $\Psi_z(R,z) = \Psi(R,z) - \Psi(R,0)$ 
governs the vertical structure of the disk, $\mathop{\mathrm{erfc}}$ is the complementary error function. In the above 
equation, $M_d$ is the disk mass, $R_d$ is the scale length and $h_z$ is the scale height.

A spherical live classical bulge is constructed from the King model \citep{King1966} 
and the corresponding distribution function (DF) is given by \citep{BT1987}
\begin{equation}
  f_{b}(E)=\left\{
    \begin{array}{ll}
      \rho_{b}(2\pi\sigma_{b}^2)^{-3/2}
      e^{(\Psi_{b0}-\Psi_{c})/\sigma_{b}^2} &\\
      \times \{e^{-(E-\Psi_{c})/\sigma_{b}^2}-1\}
        & \mbox{if} \; E < \Psi_{c},\\
      0  \quad &\mbox{otherwise}.
    \end{array} \right.
\end{equation}

\noindent Here, the bulge is specified by three parameters, namely the cut-off 
potential ($\Psi_c$ which determines the bulge tidal radius), central bulge density ($\rho_b$) and 
central bulge velocity dispersion ($\sigma_b$). The gravitational potential at the centre of the bulge
is measured by $\Psi_{b0}$.

An axisymmetric live dark matter halo is constructed using the distribution function of 
a lowered Evans model \citep{Evans1993} and is given as
 
\begin{equation}
  f_{dm}(E,L_z^2)=\left\{
    \begin{array}{ll}
     [(A L_z^2 + B)e^{-E/\sigma_h^2} + C]\\
     \times (e^{-E/\sigma_h^2} -1) & \mbox{if} \; E<0,\\
      0 \quad  &\mbox{otherwise}.
    \end{array} \right.
\end{equation}

\noindent 

\noindent The halo is parameterized by a potential depth ($\Psi_0$), velocity ($\sigma_h$) and density scales ($\rho_1$),
 a core radius $R_c$ and the flattening parameter $q$. The factors $A, B,$ and $C$ are functions of these 
parameters (see \citealp{KD1995} and references therein). The halo has a tidal radius specified by $E=0$.  

\begin{figure}
\rotatebox{270}{\includegraphics[height=7.5 cm]{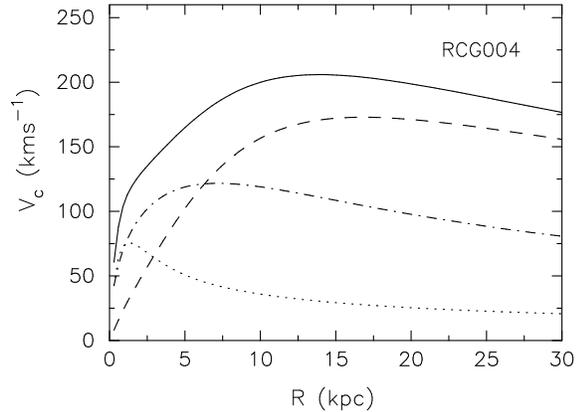}}
\caption{Initial circular velocity curve for the model galaxy. Solid line represents the total circular velocity
. Dotted line is for the bulge, dashed line for the dark matter halo and dashed-dot-dash for the disc.\label{fig:rotc}}
\end{figure}

The total mass and the outer radii of both the bulge and halo are calculated
in an iterative procedure. The potential is computed self-consistently by solving
the Poisson equation for the combined three component system in an iterative fashion.
First, the densities for the bulge and halo are obtained from their respective 
distribution function and then the disk density is added to it and the corresponding
potential for the combined mass distribution is used as a starting point for carrying out
the next iteration. We use a maximum of $l=10$ in the potential harmonic expansion
and the iteration is continued until the outer radii for the bulge and halo are 
unchanged between successive iterations. The outer radii of the bulge and the halo correspond to
the respective tidal radii.  The masses of the bulge
and halo correspond to the total mass enclosed within their respective outer radii computed by
integrating the density profiles.

In this paper, we present the analysis of a particular galaxy model hosting a low mass
classical bulge. For historical reasons, we call this model RCG004. 
The circular velocity curve for the model is presented in Fig~\ref{fig:rotc}.  
The length, mass and velocity units for this model are given by 
$L=4.0$ kpc, $M = 2.33 \times 10^{10} M_{\odot}$ and $V = 157$ kms$^{-1}$. 
The disk outer radius ($R_{out}$) is fixed at about $6.5 R_d$ and a truncation width 
$\sim 0.3 R_d$ is adopted within which the disk density smoothly decreases to zero at 
the outer radius. The disk scale length ($R_d$) is fixed at $4.0$ kpc and the scale height
is $42$ pc, the disk mass $M_d = 4.5 \times 10^{10} M_{\odot}$.  
The central value of the radial velocity dispersion is $78.5$ kms$^{-1}$. The Toomre Q profile is nearly 
flat in the radial range $0.5$ to $5$ scale lengths while it increases on either side of the disc. 
The Q value at the disc half mass radius is $1.4$.
The bulge mass $M_b = 3 \times 10^9 M_{\odot}$. In table~\ref{paratab}, we quote the
outer radius for the classical bulge (denoted by $R_b$) in our galaxy model.
The halo has a flattening of $q=0.8$ and a core radius $R_c =0.25$ kpc and a mass of 
$M_h = 1.82 \times 10^{11} M_{\odot}$ within about $60$ kpc.   
 
\begin{figure}
\rotatebox{0}{\includegraphics[height=7.5 cm]{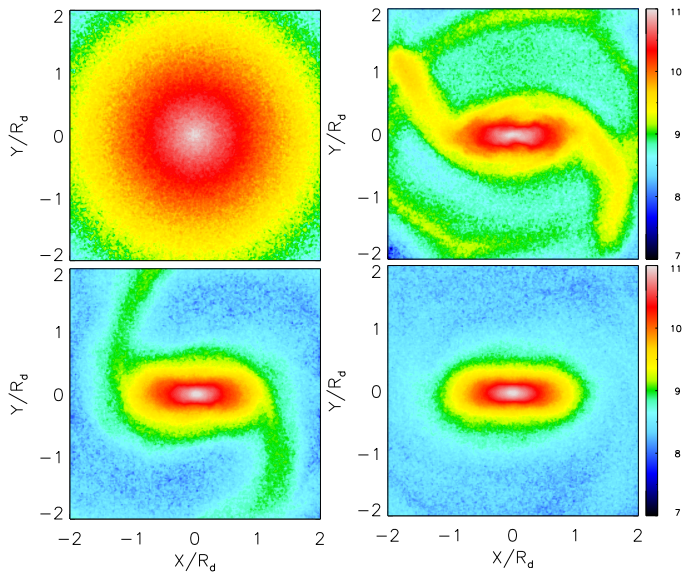}}
\caption{Surface density maps for the disk particles alone. Top left: density map at T=0 Gyr, 
showing the axisymmetric disk. Top right: Same at T=0.56 Gyr. Bottom left: at T=1.1 Gyr 
and Bottom right: at T=2.1 Gyr.  \label{fig:Ddenmaps}}
\end{figure}

We evolve the galaxy model in isolation to examine the evolution of the bulge shape, morphology and kinematics.
The simulation is performed using the Gadget code \citep{Springeletal2001} which uses a variant of the leapfrog 
method for the time integration. The forces between the particles are calculated using the Barnes \& Hut (BH) 
tree with some modification \citep{Springeletal2001} with a tolerance parameter $\theta_{tol} = 0.7$. 
The integration time step is $\sim 0.4$ Myr and the model is evolved for $2.2$ Gyr. For reference, the 
orbital time at the disk half mass radius is $\sim 296$ Myr.
A total of $1.0 \times 10^7$ particles is used to simulate the model galaxy. The softening lengths for the disk, 
bulge and halo particles are $12 $, $40$ and $36$ pc respectively. The masses for the disk, bulge and halo particles are 
$1.2 \times 10^4 M_{\odot}$, $0.3 \times 10^4 M_{\odot}$ and $3.6 \times 10^4 M_{\odot}$ respectively.
To examine the effect of unequal softenings, we have re-run the simulation with new softening parameters
 as prescribed by \cite{McMillan2007}. We note that with the new softenings, the bar growth is delayed by
$\sim 90$ Myr while the main results remain unchanged. The total energy is conserved within 0.2\% 
till the end of the simulation. The total angular momentum is conserved within 3\% at 2.2 Gyr for both the runs
having different softening parameters.   

\begin{table}
\caption[ ]{Initial parameters for the model galaxy.}
\begin{flushleft}
\begin{tabular}{cccccccc}  \hline 
Galaxy    & $Q$   & ${B}/{D}$  & ${B}/{T}$  &$\sigma_{b0}$  &  $R_b$ \\
model &              &     &    &(kms$^{-1})$  & (kpc) & &   \\
\hline

RCG004 & 1.40 & 0.0666 & 0.01306 & 65.0  & 6.08 \\
 
\hline
\end{tabular}
\end{flushleft}
{B/D is the bulge-to-disk mass ratio, B/T is the bulge-to-total (including dark halo mass) mass ratio,
$\sigma_{b0}$ is the bulge central velocity dispersion , and finally $R_b$ is the outer radius of the bulge. }
\label{paratab}
\end{table}


\begin{figure}
\rotatebox{-90}{\includegraphics[height=8.5 cm]{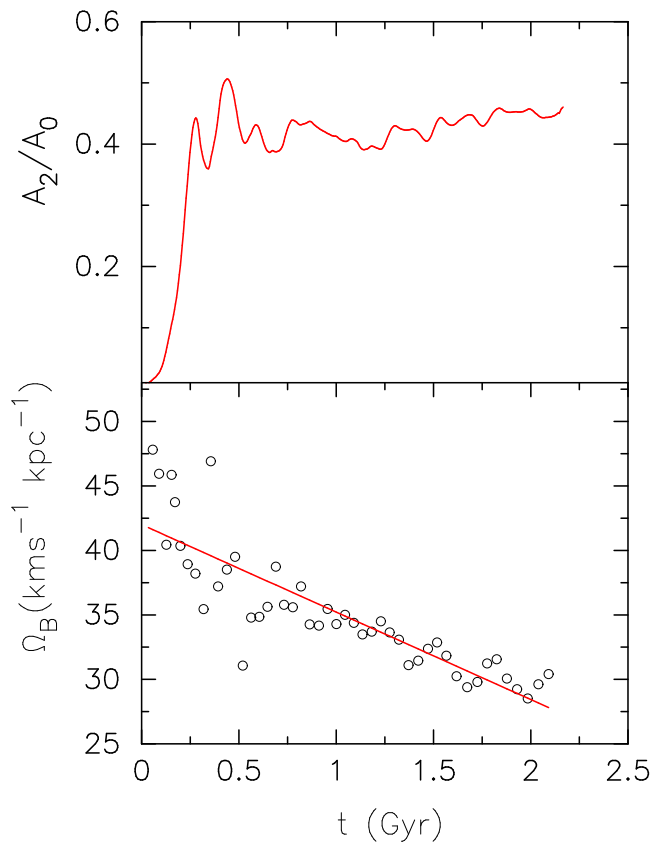}}
\caption{Time evolution of the bar amplitude and the pattern speed
 ($\Omega_B$). Red solid line is the result of a linear regression analysis done on 
the measured pattern speed values from the {\it N}-body snapshots.}
\label{fig:barAMP}
\end{figure}

\section{bar and boxy bulge}
\label{sec:barevolution}
Although it is not clearly understood how bars are formed in real galaxies, swing amplification 
\citep{Toomre1981} plays a significant role in making an initially axisymmetric, equilibrium model of 
a disk galaxy bar unstable \citep{Sellwood1981}. Once formed, {\it N}-body bars are found to be long-lived, 
dominate the disk dynamics, and are responsible for driving secular evolution processes in the galaxy 
\citep{SellwoodWilkinson1993}. Fig.~\ref{fig:Ddenmaps} depicts the formation and evolution
of the bar from the initially axisymmetric disk. Strong two-armed spirals are also formed along with the bar 
and last till 1.1 Gyr in our simulation. The ring-like structure at $T=0.56$ Gyr is intersecting the spiral 
arms indicating that it is probably not real. Such a ring-like feature arises because of the galaxy model not 
being in perfect equilibrium. 

In the upper panel of Fig.~\ref{fig:barAMP}, we show the time evolution 
of the bar amplitude measured as the maximum of $m=2$ Fourier coefficient ($A_2$) of the density perturbation 
normalized to the unperturbed axisymmetric component ($A_0$). The bar reaches its first peak in amplitude 
at $0.28$ Gyr and the second peak at $0.44$ Gyr. The $m=1$ vertical Fourier mode ($|A_{1,z}|$) in 
the $r-z$ plane corotating with the bar pattern speed shows that the disk is undergoing buckling instability from 
$\sim 0.39 - 0.6$ Gyr and strong buckling occurs around $0.6$ Gyr. 

Based on the nature of the growth curve, bars are classified into two broad categories, type-I and type-II. Type-I bars 
are strong, grow within a few orbital time scales and nearly saturate in amplitude, whereas type-II bars are weak, 
the growth time scale is very long (typically, a secular evolution time scale) and show no sign of saturation 
\citep{Sahaetal2010}. The bar in our model is a type-I bar (e.g., Fig.~\ref{fig:barAMP}). 

\begin{figure*}{}
  \newlength{\figwidth}
  \setlength{\figwidth}{\textwidth}
  \addtolength{\figwidth}{-\columnsep}
  \setlength{\figwidth}{0.5\figwidth}
  
  \begin{minipage}[t]{\figwidth}
    \mbox{}
    \vskip -1pt
    \centerline{\includegraphics[width=1.85\linewidth,angle=0]{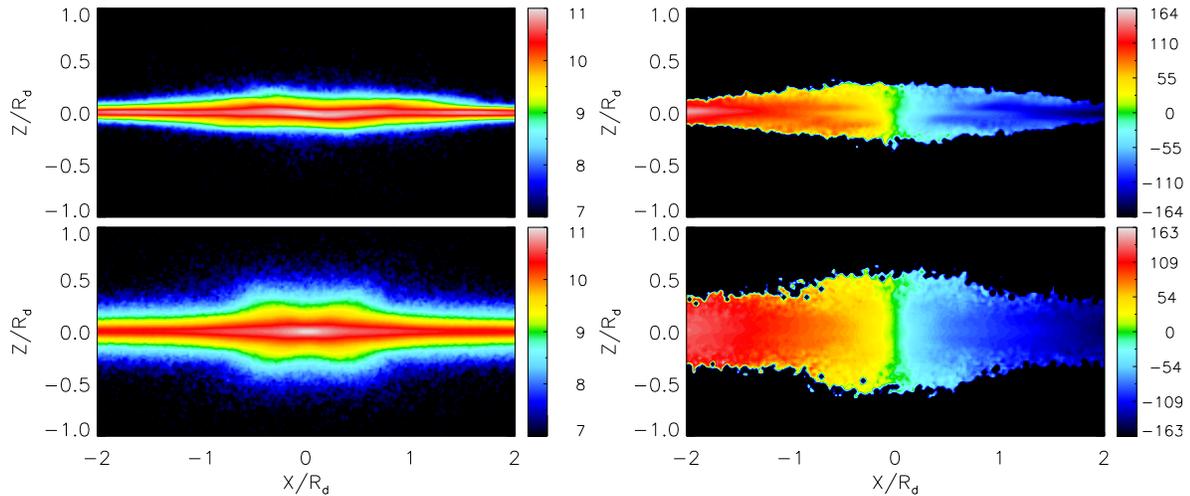}}
  \end{minipage}
  \hfill
\caption{Edge-on surface density (left) and velocity maps (right) for the disk particles alone
 at two different epochs during the secular evolution. From top to bottom, the 
panels are taken at $T= 0.56$ and $2.1$ Gyr. 
 \label{fig:boxyB}}
\end{figure*}

The bottom panel of Fig.~\ref{fig:barAMP} shows the evolution of the bar pattern speed. 
The pattern speed of the bar decreases over time in our simulation, primarily 
because of the dynamical friction \citep{TremaineWeinberg1984, Weinberg1985} against the dark matter halo. 
A detailed account on the bar's pattern speed decrease and its dependence on various dark matter 
halo properties e.g. halo angular momentum, orbital anisotropy, central concentration can be found in 
\cite{DebattistaSellwood2000}. Using a linear regression analysis on the simulation data, we find
the half-life, $T_{1/2}$, (the time period over which the bar pattern speed would decay to half its initial value) 
of the rotating bar to be $\sim 3.09$ Gyr. This indicates that the rate of angular momentum transfer from the bar 
is rather slow in our simulation; for an in-depth analysis on the bar slow down, readers are referred to 
\cite{Weinberg1985,WeinbergKatz2007a}. 

As the bar grows stronger, its self-gravity increases and it goes through the well-known buckling instability 
\citep{CombesSanders1981,PfennigerNorman1990, Rahaetal1991, MV2004} following which the bar transforms into a 
boxy/peanut bulge. In Fig.~\ref{fig:boxyB}, we present the surface density (left panels) and 
velocity field (right panels) for the boxy bulge seen edge-on; i.e., only disk particles are shown. The cylindrical 
rotation is evident. The final boxy bulge contains approximately $33\%$ of the disk mass including the inner 
barred disk component.
Note that the density drops off sharply along the vertical direction in the boxy bulge region. In 
Section~\ref{sec:bulgestructure}, we will compare the structure and kinematics of the boxy bulge in 
Fig.~\ref{fig:boxyB} formed in our simulation with the classical bulge undergoing the bar driven secular evolution. 

\section{Angular momentum transfer to the classical bulge}
\label{sec:Lzexchange}
\begin{figure}
\rotatebox{-90}{\includegraphics[height=8.5 cm]{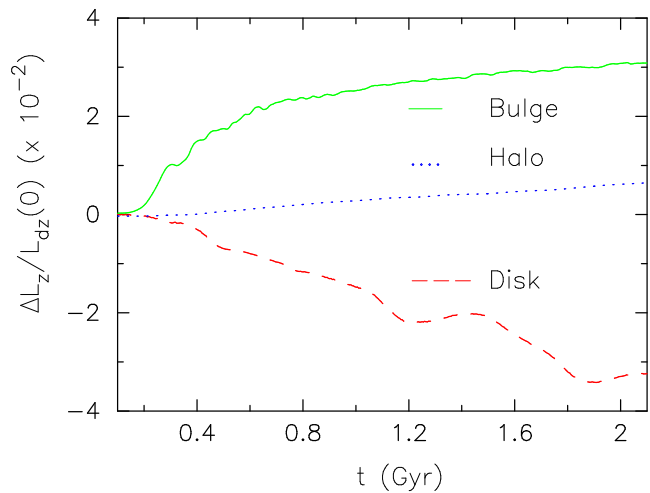}}
\caption{Evolution of the specific angular momentum of the bulge (green), disk (red) and halo (blue) 
components in our model. Along the y-axis plotted are the specific angular momentum minus its value at $T=0$ 
normalized by the disk angular momentum ($L_{dz}(0)$) at $T=0$.}
\label{fig:spLz}
\end{figure}

We compute the specific angular momentum for each species e.g., disk, bulge and halo particles
in our simulation and re-confirm the already established fact that the inner regions of the disk loose 
angular momentum through the bar. While a significant fraction of the total angular momentum emitted 
by the bar is absorbed by the surrounding dark matter halo, the angular momentum gained by the bulge 
is non-negligible. In Fig.~\ref{fig:spLz}, we show the angular momentum 
transfer amongst the disk, bulge and halo components in our model. The total angular 
momentum is conserved within $3\%$ at the end of 2.2 Gyr in our simulation.   
Initially both the bulge and halo have zero net angular momentum i.e., they start as 
non-rotating objects. Note that the rate of gain of angular momentum 
by the classical bulge particles nearly saturates towards the end of the simulation and closely 
follows the growth of the bar (see Fig.~\ref{fig:barAMP}).
Using orbital spectral analysis, we show below that the gain of angular momentum
by the bulge occurs primarily through resonances \citep[see also] [] {HernquistWeinberg1992}.

\begin{figure*}
\begin{centering}
\includegraphics[angle=-90,scale=0.62]{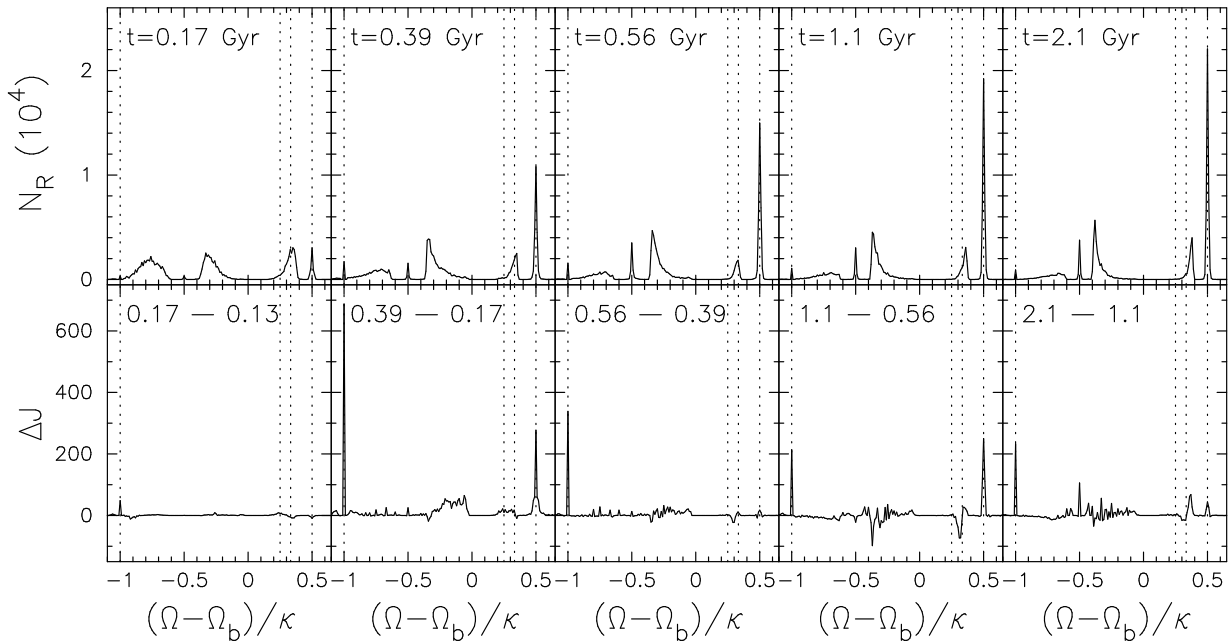}
\end{centering}
\caption{The top panels show the distribution of particles with frequency $(\Omega-\Omega_b)/\kappa$ 
at five different times throughout the evolution of the bulge. The lower panels show the gain of angular 
momentum of the selected particles with respect to the previous time, as indicated on the top of the panel. 
The vertical dotted lines in the figure indicate the most important resonances, -1:1, 2:1, 3:1 and 4:1. 
Note that from time $1.1$~Gyr to $2.1$~Gyr the particles trapped at around $(\Omega-\Omega_b)/\kappa$=0.5 
gain less angular momentum than earlier.}
\label{fig:resonances}
\end{figure*}

The simulation presented here shows an increase in the bulge rotation velocity 
(Section~\ref{sec:CBrotation}), and a corresponding increase in bulge angular 
momentum. The transfer of angular momentum from the bar to the bulge depends 
strongly on the pattern speed which sets the resonance locations. It is important 
to note that if the resonances are sparsely populated because of lack of particles 
in the simulation, the angular momentum transfer will be inefficient \citep{Weinberg1985, 
WeinbergKatz2007b}. In our case, we have a total of $10^7$ particles with $10^6$ particles 
in the classical bulge. So we can test whether angular momentum transfer through 
resonant interaction, is the mechanism for the angular momentum gain of the classical bulge. 
Here, we quantify its effect by using an orbital spectral analysis method described in 
\citet{MV2006} based on that presented in \citet{BinneySpergel1982}. In previous works, 
this method was applied to halo resonant orbits \citep{Athanassoula2003,MV2006,DubinskiBerentzen2009} 
to understand the bar-halo interaction. We apply it to the classical bulge particles to find out 
how many of them are trapped in resonances and what is the corresponding gain of 
angular momentum. The potential is extracted from the {\it N}-body simulation at 
different snapshots using the grid code provided by \cite{SellwoodValluri1997} and then 
frozen to compute the orbits. We randomly select $100000$ ($10 \%$) particles out 
of 1 million in the bulge and compute their corresponding orbits. We calculate the 
azimuthal and radial epicyclic frequencies $\Omega$ and $\kappa$ respectively for each 
of the orbits by Fourier analysis.

We present the results of this orbital spectral analysis in
Fig.~\ref{fig:resonances} at $5$ different epochs. On the top panels, we present the
classification of classical bulge particles by their frequency ratio
$\eta=(\Omega-\Omega_B)/\kappa$. The bar has an irregular evolution and this can be
seen in the time sequence of the top panel in Fig. \ref{fig:resonances}. Initially,
the bulge particles are distributed half co-rotating with the disk and half
counter-rotating. Therefore, when we study the orbital distribution in the very early
stage of the bar growth at $T=0.17$~Gyr, there still is an almost symmetric
distribution. When the bar is already formed, right after reaching the
maximum, at $T=0.39$~Gyr, many particles have been trapped around the 2:1 resonance.
Taking a careful look at these orbits, we have checked that they are of x1-type.
There is also a considerable group of particles with $\eta \in (-0.4,0.)$; a look to
the orbits allows to identify them as mainly stochastic trajectories. In the lower
panels, we show the angular momentum gain by the particles at each $\eta$ during the growth 
and evolution of the bar. There is a considerable gain of angular momentum by three main groups in our
diagram (Fig.~\ref{fig:resonances}, second bottom panel). The main gain of angular momentum comes from
those particles at resonance with  $\eta = -1$, corresponding to
particles orbiting around the Lagrangian points. Since these particles are at
negative frequency it means that they are counter-rotating with the bar. Another
gaining group corresponds to the particles with $\eta \in
(-0.4,0.)$, the stochastic group. By gaining angular momentum their (counter) rotation decreases. Amongst 
the low order resonances, the important gaining group is around the ILR ($\eta=0.5$). 
We can conclude that at this stage of evolution, which is very rapid, the main transfer of angular momentum
occurs through resonant and stochastic orbits.

\begin{figure*}{}
  \setlength{\figwidth}{\textwidth}
  \addtolength{\figwidth}{-\columnsep}
  \setlength{\figwidth}{0.5\figwidth}
  
  \begin{minipage}[t]{\figwidth}
    \mbox{}
    \vskip -1pt
    \centerline{\includegraphics[width=0.85\linewidth,angle=0]{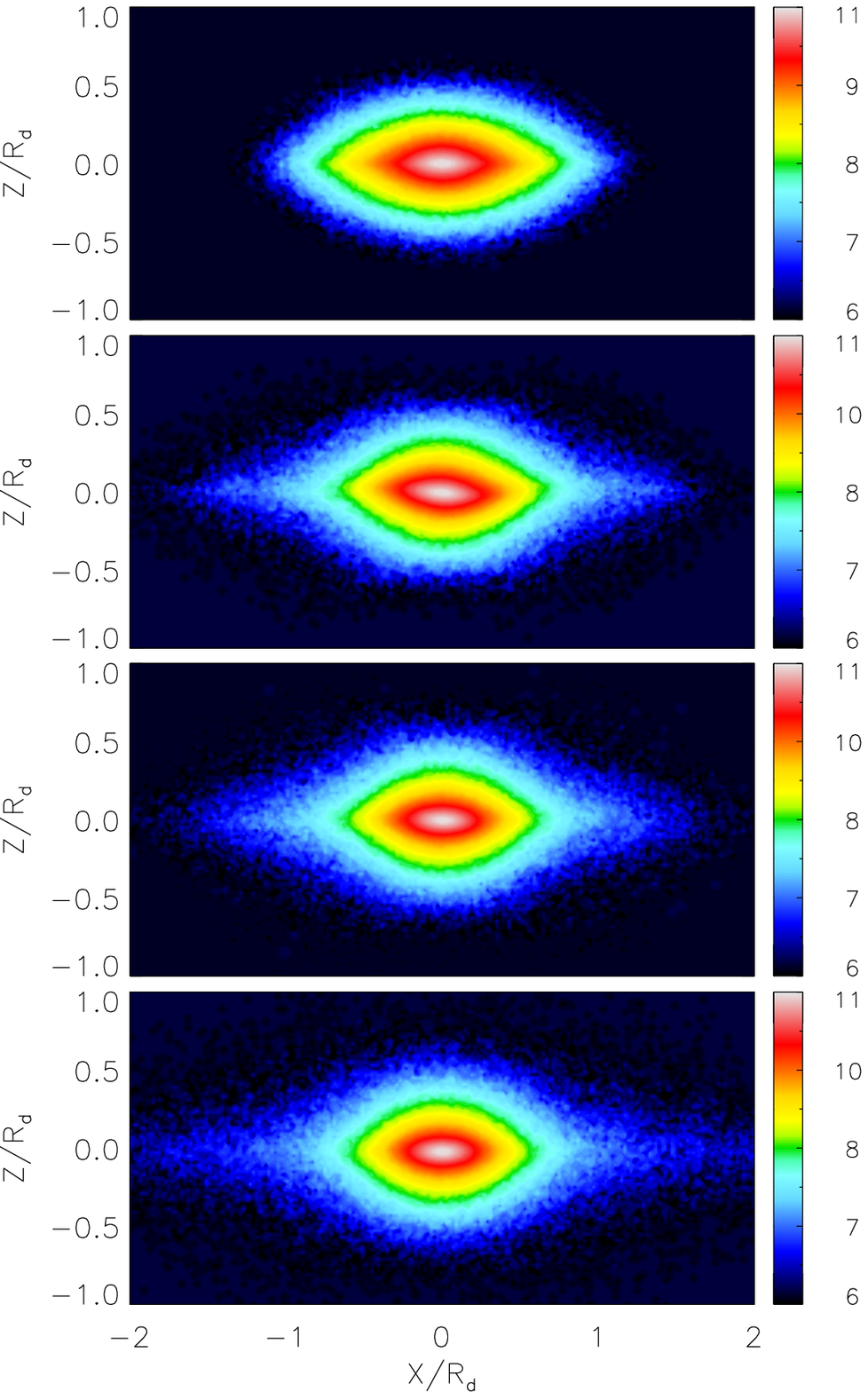}}
  \end{minipage}
  \hfill
  \begin{minipage}[t]{\figwidth}
    \mbox{}
    \vskip -1pt
    \centerline{\includegraphics[width=0.85\linewidth,angle=0]{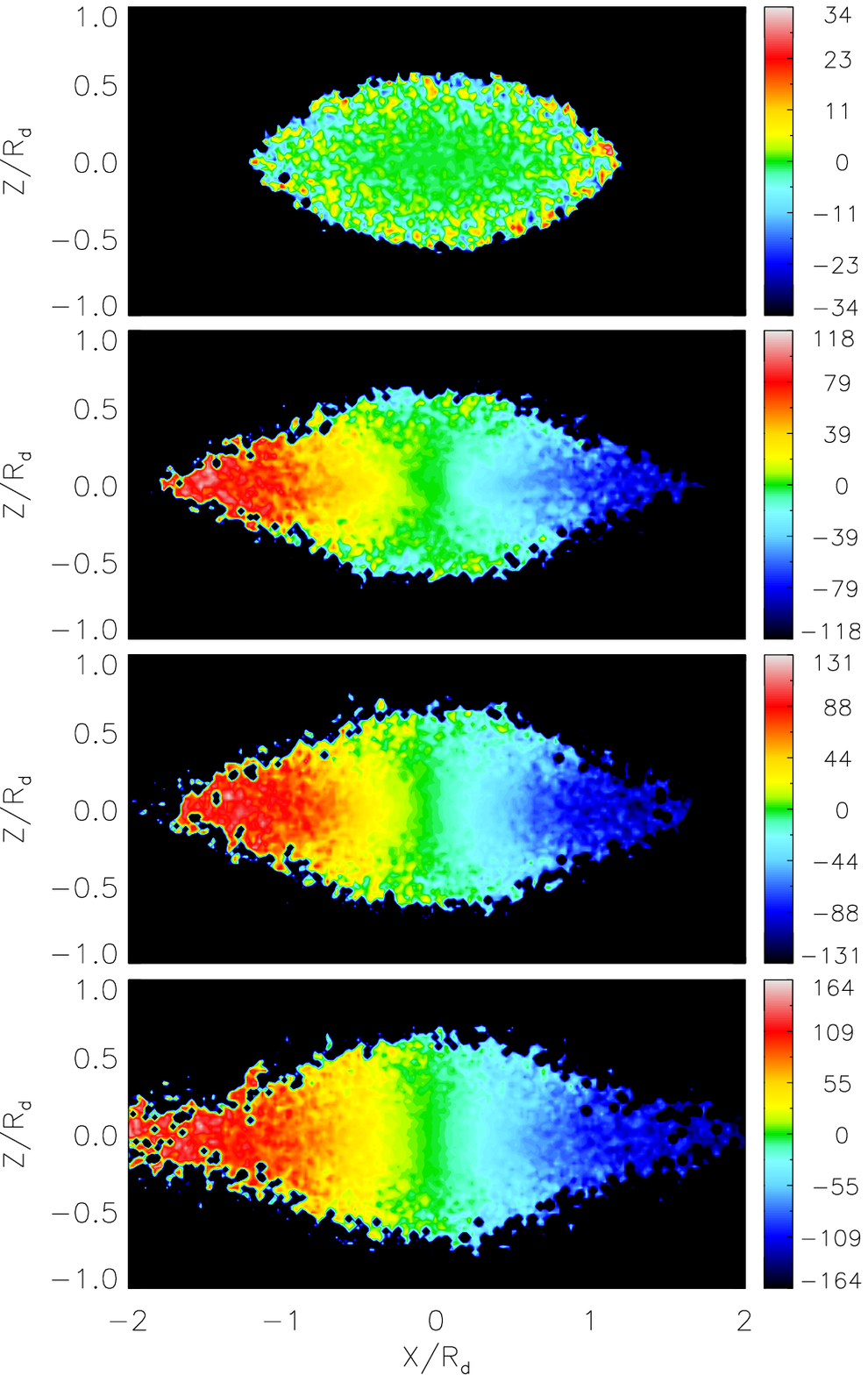}}
  \end{minipage}

    \caption{Edge-on surface density  and velocity maps for the bulge particles alone
 at four different epochs during the secular evolution. From top to bottom, the 
panels are taken at T=0, 0.56, 1.1 and 2.1 Gyr. The left panels show the surface densities 
and the right panels the velocity fields. Initially the bulge is non-rotating and flattened by 
the disk potential. Similar maps for the boxy-bulge are shown in Fig.~\ref{fig:boxyB}. 
 \label{fig:Bdenmaps}}
\end{figure*}

During the period of evolution between $T = 0.56-0.39$~Gyr, the bar goes through the
buckling event, therefore there is still some trapping of particles around the 2:1
resonance. Although there not much gain or loss of angular momentum, some 
angular momenta are gained through $ \eta = -1$ resonance during this period.
Notice that in the upper row of panels, the number of bulge particles trapped at
the bar's ILR (2:1) is increasing. In the forth panel, at $T=1.1$~Gyr, we have $36\%$ of particles trapped
around the 2:1 resonance. At this time, the main gain of angular momentum comes from
resonances at the ILR ($\eta = 0.5$), and at $\eta = -1,-2$. The particles at the OLR
($\eta = -0.5$) and those with $\eta \in (-0.4,0.)$ are losing angular momentum. At
$T = 2.1$~Gyr the classical bulge is still gaining angular momentum through resonances
at ILR and $\eta = -1$ corresponding to the Lagrange points. Note that some of the
angular momentum gain is also coming from the OLR. Although the number of particles
trapped at the ILR is gradually increasing over time, their angular momentum gain
does not follow accordingly. By comparing the last two panels (upper and lower), it is
evident that the particles trapped at the ILR are now hardly gaining angular momentum.
It is plausible that the inner bar-like structure in the classical bulge (see section~\ref{sec:bulge-bar})
is giving away angular momentum to the outer parts of bulge and perhaps to disk and halo.   

The gain of angular momentum by the bulge can thus be explained by resonances during
the slow secular evolution of the bar, and by resonances together with stochastic
orbits in the dynamical stage. During the dynamical phase, $T = 0.56 - 0.17$~Gyr, the net gain 
of angular momentum (computed by adding up the averaged angular momentum of each orbit) is 
$3$ times larger than that gained in the relatively quiet secular phase ($T = 2.1 - 0.56$~Gyr).
While approximately $70\%$ of the net angular momentum gain comes from the resonances, stochastic orbits 
contribute to $\sim 30 \%$ of the net angular momentum gained during the dynamical phase.  

Previous studies of angular momentum transfer to the live dark matter halo 
(e.g., \citet{Athanassoula2003, MV2006, DubinskiBerentzen2009}) have found important contribution
 from corotation, outer Lindblad resonance (OLR) and higher order resonances. 
By contrast, in the case of a small classical bulge as studied here, the OLR and corotation
have not played any significant role in the gain of angular momentum as shown above.
This is most probably due to the fact that the size
of the bulge in our simulation is much smaller than the typical size of the dark
matter halo; the bulge half-mass radius ($R_{1/2}^b = 0.21 R_d$ and $0.225 R_d$ at
$T=1.1$ and $2.1$ Gyr respectively) is shorter than the bar size ($ R_{bar} = 0.987 R_d$
and $2.01 R_d$ at those times) in our simulation. The bar size is measured from the phase 
angle of the bar \citep{Athanamisi2002}. The phase angle of the bar (i.e., the $m=2$ Fourier component 
of the disk surface density) remains approximately constant upto a certain radius and starts 
varying beyond that. We measure the length of the bar as the radius at which the phase angle 
of the bar starts deviating from the contant value.
We also note that the corotation resonance ($R_{cr} = 0.994, 1.12$ and $1.30 R_d$ at $T =
0.56, 1.1$ and $2.1$ Gyr respectively) of the bar is clearly outside the radius
confining most of the bulge particles. The ratio of $R_{cr}/R_{bar}$ lies
between $1 - 1.4$ at times mentioned in Fig.~\ref{fig:resonances}. As shown in
Fig.~\ref{fig:resonances}, more and more particles in the classical bulge are 
trapped at the bar's ILR as time progesses. During the slow secular evolution, we can affirm that the
mechanism acting in our system is the transfer of angular momentum through
resonances. On the other hand, during the rapid dynamical evolution, resonant as well 
as stochastic orbits played an important role in transferring a significant fraction
of the net angular momentum to the classical bulge.
This angular momentum transfer and the subsequent change of the orbital structure of the
classical bulge are indeed responsible for the transformation of the classical bulge
as described below. 
  
\section{Evolution of the classical bulge - structure and kinematics}
\label{sec:bulgestructure}
From Sections~\ref{sec:barbulgeinteraction} and \ref{sec:Lzexchange}, we learn that a classical bulge can 
absorb a non-negligible fraction of the total angular momentum emitted by the bar through resonant interaction. 
The angular momentum gained by the bulge (being a smaller mass object than the dark matter halo) has a profound 
effect on its structure, kinematics and dynamics. The result of the bar-bulge interaction in 
our simulation is the {\it transformation of an initially non-rotating low mass classical bulge 
into a highly rotating triaxial one}. Below we describe, in considerable detail, various 
diagnostics which show that this is indeed true. 

\begin{figure}
\rotatebox{-90}{\includegraphics[height=8.5 cm]{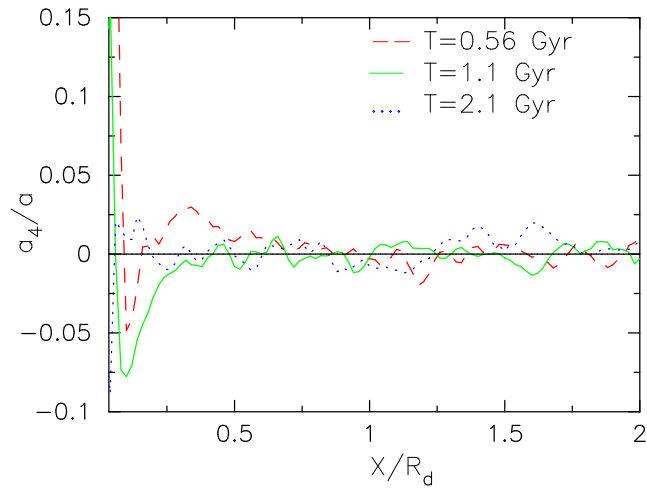}}
\caption{Normalized radial profile for the $a_4$ coefficient for the classical bulge alone at three different times.}
\label{fig:a4bulge}
\end{figure}

\subsection{Surface brightness}
\label{sec:CBsurfb}
In the left panels of Fig.~\ref{fig:Bdenmaps}, we show the surface density maps for the classical bulge
 (viewed edge-on) at four different epochs during the evolution. The classical bulge is shown edge-on 
($i = 90\,^{\circ}$) such that the major axis is along $X$-axis and the minor axis along $Z$-axis. Initially the bulge 
is isotropic and flattened by the strong gravity of the disk potential. At later phases of evolution, the inner regions of 
the bulge become rounder and the outer parts become disky. 

In order to understand the structure of the classical bulge 
more quantitatively, we have also performed an isophotal analysis using the IRAF ellipse task on a set of edge-on 
images of the bulge including the ones presented in Fig.~\ref{fig:Bdenmaps}, and compute the fourth-order Fourier cosine 
coefficient $a_4/a$ normalized to the semi-major axis $a$ at which the ellipse was fit.   
Fig.~\ref{fig:a4bulge} shows the normalized $a_4$ prodiles at three different epochs in the simulation. The values of $a_4/a$ determine 
the degree of boxiness or diskiness \citep{NietoBender89}, with $a_4/a < 0$ denoting a boxy isophote, $a_4/a > 0$ a disky 
isophote, and $a_4/a \sim 0$ means elliptical or round isophote. The inner region ($ X/R_d \le 0.2$) of the classical bulge
becomes mildly boxy at $T = 0.56$~Gyr and the boxiness increases at $1.1$~Gyr as can be seen from Fig.~\ref{fig:a4bulge} and Fig.~\ref{fig:Bdenmaps}. On the other hand, at $T = 2.1$~Gyr, $a_4/a > 0$ in the outer parts of the bulge indicating disky isophotes.
Recall that the disk has a boxy bulge formed as a result of the bar buckling instability 
as shown in Fig.~\ref{fig:boxyB}. In order to compare properties of the classical bulge and boxy bulge, we have measured the $a_4/a$ parameter of the 
boxy bulge in a similar fashion as outlined above. It is found that at $T=1.1$~Gyr, $a_4/a$ of the boxy bulge is negative inside $X/R_d < 1.0$ and 
its maximum value is about twice that of the classical bulge.
From the minor axis density profiles calculated separately for the classical bulge and the boxy bulge region, we find that initially 
the classical bulge extends further above the disk midplane, and its central surface density is $2.4$ times lower than that of the disk. 
As the disk goes through the buckling instability, the particles settle into the 3D boxy bulge. We find that at $T=1.1$~Gyr the boxy bulge
is more concentrated toward the disk midplane ($z = 0$) and its midplane surface density is $\sim 3$ times higher than that of the classical bulge. At this time, 
the density of classical bulge above $z = 0.17 R_d$ is higher than that of the boxy bulge, and the classical bulge extends further. 
However, at $T=2.1$~Gyr, the boxy bulge dominates over the classical bulge for $z \le 0.39 R_d$ and above this height, their density profiles 
are comparable.  

\begin{figure}
\rotatebox{-90}{\includegraphics[height=8.5 cm]{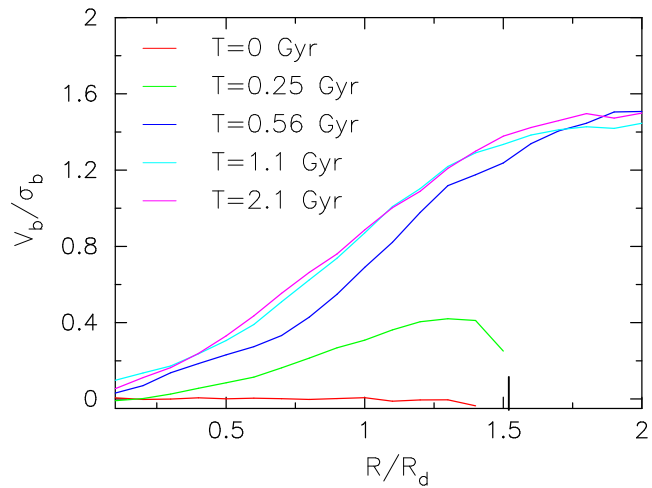}}
\caption{Radial variation of the bulge rotational velocity normalized to the average velocity dispersion in the central 
region, for five snapshots from T=0 to T=2.1 Gyr. The long tick mark on the x-axis denotes the initial 
value of $R_b$, see Table~\ref{paratab}.}
\label{fig:vcbulge}
\end{figure}

\begin{figure}
\rotatebox{-90}{\includegraphics[height=8.5 cm]{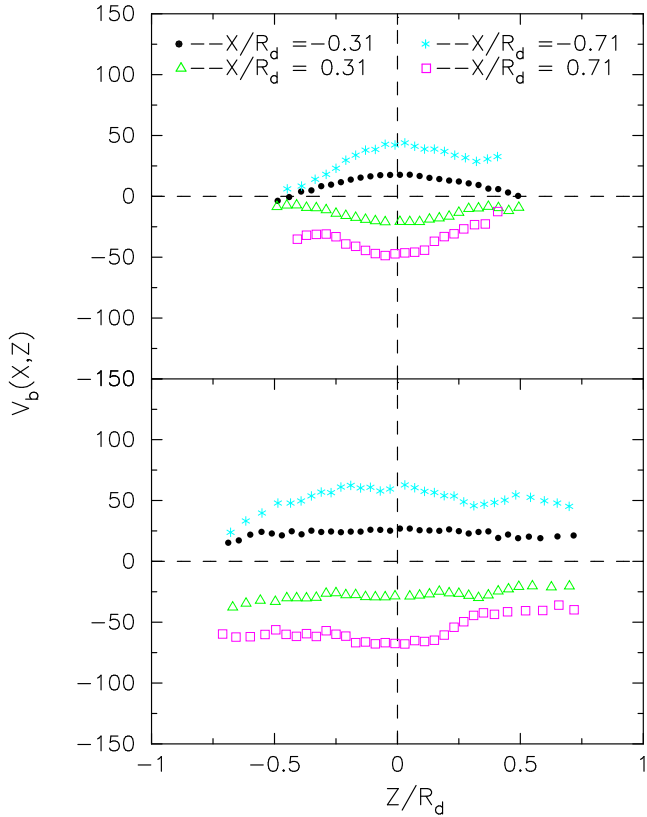}}
\caption{Parallel minor axis velocity profiles of the bulge at T=0.56 Gyr (upper panel) and T=2.1 Gyr (lower panel). 
The upper panel shows no clear signature of cylindrical rotation. But at later stages of the evolution, 
cylindrical rotation develops in the inner region of the bulge, as indicated by the parallel shapes of the 
velocity profiles in the lower panel.}
\label{fig:vcB_minor}
\end{figure}

\begin{figure}
\rotatebox{-90}{\includegraphics[height=8.5 cm]{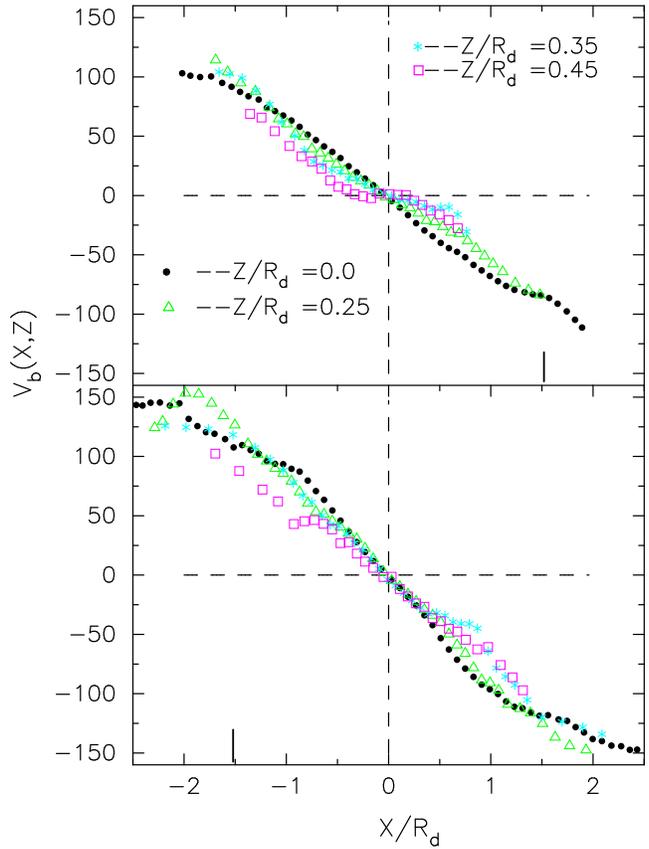}}
\caption{Major axis velocity profiles of the bulge at T=0.56 Gyr (upper panel) and T=2.1 Gyr (lower panel). 
The upper panel shows no clear signature of cylindrical rotation. But at later stages of the evolution, cylindrical 
rotation develops in the inner region of the bulge. The long tick mark on the x-axis denotes the initial value 
of $R_b$, see Table~\ref{paratab}.}
\label{fig:vcB_major}
\end{figure}

\subsection{Rotational properties}
\label{sec:CBrotation}
The influx of angular momentum to the initially non-rotating bulge enforces the 
bulge particles to have a net rotational motion. In Fig.~\ref{fig:vcbulge}, we show the radial profiles of the 
rotational velocity normalized to the average velocity dispersion in the central region ($ R \le R_{1/2}^b$) of 
the classical bulge at different epochs during the evolution. The rotational velocity profiles remain 
nearly unchanged at later stages of evolution when the rate of angular momentum gain by the bulge also 
nearly saturates as can be seen from Fig.~\ref{fig:spLz} and Fig.~\ref{fig:resonances}.

To illustrate the evolution further, we present four velocity maps of the classical bulge on the right panels 
of Fig.~\ref{fig:Bdenmaps}. During the initial phases of secular evolution, the 
angular momentum gained by the bulge particles is primarily converted into streaming motion and the classical 
bulge starts rotating around the z-axis, with gradients in the streaming velocity both along the radial and 
vertical directions. Note that in the barred potential, the classical bulge is no longer axisymmetric 
and its inner regions becomes moderately boxy (see Sections~\ref{sec:CBsurfb} and \ref{sec:bulge-bar}). 
This could be a signature of a thick bar formed inside the classical bulge 
(see Section~\ref{sec:bulge-bar}). Indeed as time progresses, mild signatures of cylindrical rotation emerge 
in the inner regions of the bulge and gradually become prominent (see Fig.~\ref{fig:Bdenmaps}).

To have a clearer picture of the velocity structure, we show parallel minor-axis (Fig.~\ref{fig:vcB_minor}) 
and major-axis (Fig.~\ref{fig:vcB_major}) velocity profiles of the classical bulge at two different epochs 
$T = 0.56$ (upper panels) and $T = 2.1$~Gyr (bottom panels). The minor-axis velocity profiles are drawn at 
two different radii ($X/R_d=0.31$ and $X/R_d=0.71$) on either side of the bulge center. At $T = 0.56$~Gyr, the 
minor axis rotation velocity decreases along the vertical direction ($\frac{d V_b}{d z} < 0$) 
indicating clearly a non-cylindrical rotation throughout the bulge (see the upper panel of Fig.~\ref{fig:vcB_minor}).
Note, the velocity profiles in the outer parts are asymmetric which is probably influenced by the on-going buckling
instability of the bar. 
The major-axis profiles are taken at four different slits (the slit positions are indicated in Fig.~\ref{fig:vcB_major})
 parallel to the major axis of the classical bulge. The major-axis velocity profiles in the upper panel of 
Fig.~\ref{fig:vcB_major}, also indicate non-cylindrical rotation throughout the classical bulge.

At later times, the inner regions of the classical bulge have developed cylindrical rotation. However, the gradient of this 
cylindrical rotation in the classical bulge is shallower than that in the boxy bulge. The minor-axis
velocity profiles in the bottom panel of Fig.~\ref{fig:vcB_minor}, show clear indication for that in the inner regions. 
The same is evident from the bottom panel of Fig.~\ref{fig:vcB_major}. 
  The major axis velocity profiles at $T = 2.1$ Gyr clearly demonstrate that the inner regions ($X/R_d < 0.4$) of 
the classical bulge rotate cylindrically while the outer regions beyond about twice the half-mass radius 
($2\times R_{1/2}^b \sim 0.45 R_d$) still maintain differential rotation both along the radial and vertical 
directions ($\frac{d V_b}{d z} < 0$). {\it So in the later stages of the secular evolution, the initially 
non-rotating classical bulge has developed a mixed rotational state with the inner region rotating 
cylindrically while the outer region rotates differentially in z.}

\subsection{The classical bulge-bar}
\label{sec:bulge-bar}
In order to achieve a deeper understanding of the complex non-linear dynamical interplay of the bar and 
the bulge, we investigate the three dimensional structure of the classical bulge using spherical harmonics analysis. 
In particular, we are looking for non-axisymmetric modes in the classical bulge which could have
been influential for producing some of the complex structure and kinematics as discussed in 
Sections~\ref{sec:CBsurfb} and \ref{sec:CBrotation}.
The outcome of the bar-bulge interaction is not only the transfer of angular momentum between the two and 
changes in kinematics thereby, but a structural transformation of the classical bulge, a prediction of which is probably
beyond the scope of the analytic/semi-analytic theories \citep{Lynden-BellKalnajs1972, TremaineWeinberg1984} 
briefly outlined in Section~\ref{sec:barbulgeinteraction}. From our analysis, it becomes clear that the 
interaction of a bar and a small classical bulge is more vigorous than that between the bar and the massive dark halo 
as discussed in section~\ref{sec:barbulgeinteraction}. The primary reason being the smaller mass and size of the bulge 
compared to the dark matter halo. 

To analyze the structural components developed in the small classical bulge after the evolution, we 
expand the full three dimensional bulge density distribution ($\rho$) in terms of spherical harmonics:

\begin{equation}    
\rho(r,\theta,\phi) = \sum_{l=0}^{\infty} \sum_{m=-l}^{l}{\rho_{lm}(r) Y_{l}^{m}(\theta,\phi)},
\end{equation}

\noindent where $r, \theta, \phi$ are the usual spherical coordinates, and the $Y_{l}^{m}$ are the spherical harmonics. 
$\rho_{lm}$ denotes the radial density function. We bin the bulge particles into spherical shells and compute $B_{lm}$ as 
function of the bin radius ($r_k$), as follows:

\begin{equation} 
B_{lm}(r_k) = N_{lm} \sum_{j} m_b P_{l}^{m}(\cos\theta_{j}) e^{im \phi_{j}},
\label{eq:blm}
\end{equation}

\noindent where $N_{lm} = ((2l+1)/2\pi)\times (l-m)!/(l+m)!$, $m_b$ is the mass of each 
bulge particle, $P_{l}^{m}$ are the Associated Legendre polynomials. The function $B_{lm}$
is directly related to the mass of each bin and thereby to $\rho_{lm}$ via the bin radius ($r_k$).
Then using the above formula (Eq.~\ref{eq:blm}), we can derive the 
radial variation of the amplitude of a particular $l, m$ mode in the bulge as 
$S_{lm}(r_k) = \sqrt{\Re{B_{lm}}^2 + \Im{B_{lm}}^2}$. The corresponding phase angle $\phi_j$ can
be used to derive the pattern speed of the $l,m$ mode.

\begin{figure}
\rotatebox{-90}{\includegraphics[height=7.5 cm]{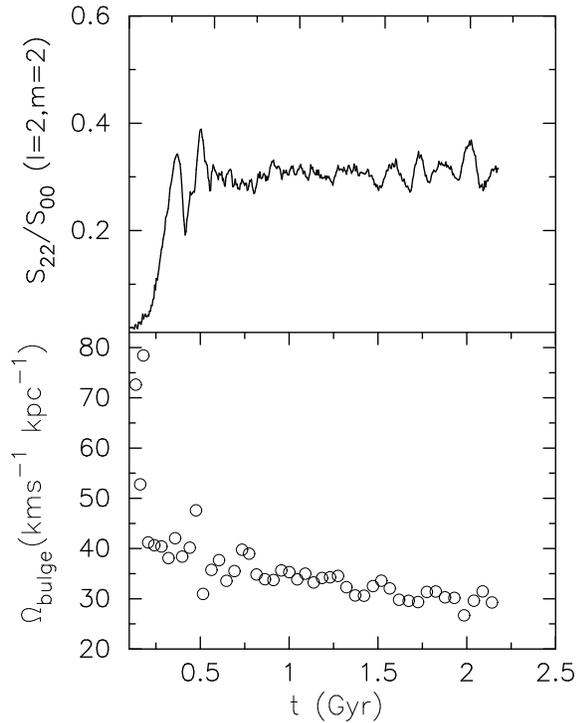}}
\caption{The strength of the classical bulge-bar (l=2,m=2 mode) and its pattern speed evolution.}
\label{fig:bulgebarlm2}
\end{figure}

In Fig.~\ref{fig:bulgebarlm2}, we show the time evolution of the amplitude of $l=2, m=2$ 
mode. The classical bulge-bar (hereafter, denoted as ClBb) is weaker than the disk bar (see Fig.~\ref{fig:barAMP} for the bar amplitude and pattern speed)
 but rotates nearly in phase with it. 
By analyzing the radial variation of the phase angles, we conclude that the physical size of the ClBb is much smaller compared to the disk bar. Initially, the 
ClBb and disk bar are not in phase, the ClBb seems to be lagging behind the disk bar by about $2^{\circ} - 4^{\circ}$ in angle. But
soon, they start rotating in-phase with each other. After about $1$ Gyr, the pattern speed of the ClBb is also nearly the
same as that of the disk bar (see Fig.~\ref{fig:bulgebarlm2}). A convenient way of viewing the dynamics of the ClBb, is to
think of it initially as a driven oscillation phenomenon where the disk bar is acting as a driver and the bar-like structure 
in the classical bulge is its forced response. Later bulge particles are trapped by the 2:1 resonance; i.e., both components populate 
the orbits in their jointly rotating potential. 
It has been shown in previous simulations 
e.g., by \cite{Holley-Bockelmannetal2005}, \cite{colinetal2006}, \cite{Athanassoula2007} that a bar-like structure also forms 
in the inner regions of the dark matter halo as a result of its interaction with the bar in the disk. These studies have shown
that such a bar in the halo is rather weak and nearly corotates with the bar in the disk.
It turns out that some of the characteristics of the ClBb are quite similar to that of the halo-bar. However, with the classical bulge being much 
less massive than the halo, the dynamical impact of the bar-like structure is much more pronounced in the classical bulge as we have already 
demonstrated above.  
Beside the transfer of energy and angular momentum between the disk-bar and the classical bulge, the stars in the classical bulge 
are also being heated during the evolution and hence the inner bulge region becomes moderately thicker and rounder (see Fig.~\ref{fig:Bdenmaps}). 
We have checked that the slow variation 
in the ellipticity of the classical bulge is consistent with the variation in the kinetic energy tensor in accordance with the tensor virial theorem.
A more detailed picture of the dynamics of the bulge hosting a bar and its observational properties will be presented in a future 
paper.

\begin{figure}
\rotatebox{-90}{\includegraphics[height=7.5 cm]{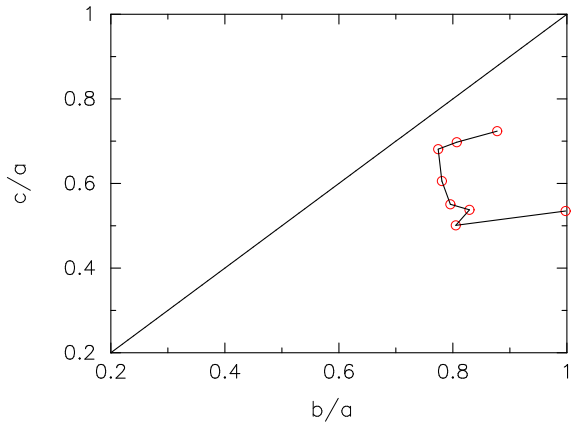}}
\caption{Time evolution of the shape of the classical bulge. Initially the bulge is flattened by the 
strong disk potential and hence oblate. At later phases during the secular evolution it becomes triaxial. 
The solid diagonal line denotes a prolate configuration. The red open circles are the measured values of
the axes ratios at $T = 0, 0.25, 0.393, 0.56, 1.1, 1.52, 1.77, 2.1$ Gyr. $b/a = 0.99$ 
and $0.877$ at $T=0$ and $2.1$~Gyr respectively.} 
\label{fig:bulgetrix}
\end{figure}

\begin{figure}
\rotatebox{-90}{\includegraphics[height=8.0 cm]{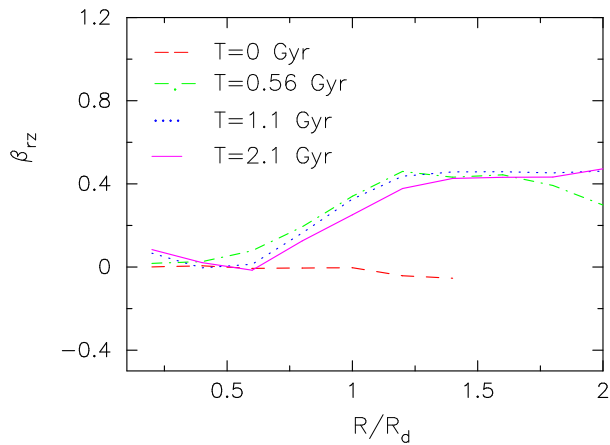}}
\caption{Radial variation of the anisotropy parameter $\beta_{rz}$ for the classical bulge. 
Beyond about $0.56$~ Gyr, the anisotropy profiles remains nearly unchanged.} 
\label{fig:bulgeaniso}
\end{figure}

\begin{figure}
\rotatebox{-90}{\includegraphics[height=8.5 cm]{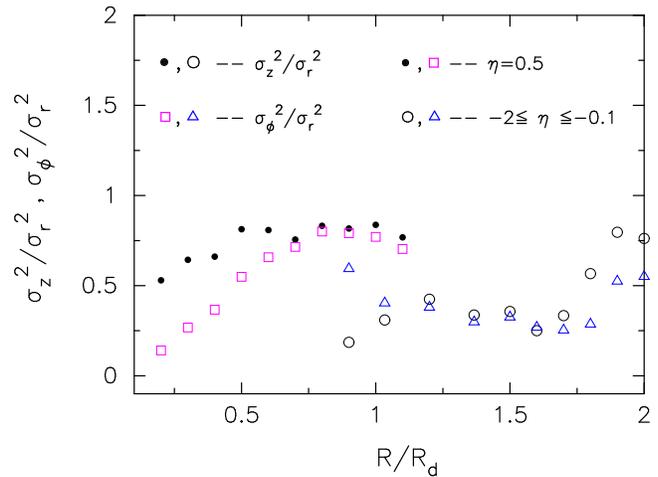}}
\caption{Radial variation of velocity dispersion ratios computed from bulge particles trapped at ILR ($\eta =0.5$) and
from a group of resonant and non-resonant particles in the frequency range $-2.0 \le \eta \le-0.1$ in the classical bulge. 
Orbits of these particles were computed at 1.1 Gyr.} 
\label{fig:orbitaniso}
\end{figure}

\subsection{Triaxiality and anisotropy}
\label{sec:Bulge triaxiality}
 From the misalignment of the photometric major axis of the disk 
and the bulge and the isophotal twists, it is inferred that many of the bulges in spiral galaxies 
are indeed triaxial \citep{Stark77,Gerhardetal1989, Bertolaetal1991, Ann1995, Mendezetal2010}. 
Here, we show the evolution of the triaxiality and velocity anisotropy in the low mass 
classical bulge in our simulation. 
The global parameter for the bulge triaxiality $T_b$ can be computed using the following 
relation \citep{Franxetal1991, Jesseitetal2005}:

\begin{equation}
T_b = \frac{1 - (b/a)^2}{1 - (c/a)^2},
\end{equation}   

\noindent where $a$, $b$, and $c$ are the semi-axes defining the shape of the classical bulge (see 
Section~\ref{sec:v-sig-epsa} for the measurement of the axis ratios). $a = b > c$ 
denotes an oblate configuration i.e. $T_b =0$, and $b = c < a$ is a prolate figure corresponding to $T_b =1$. 
$a \neq b \neq c$ defines a triaxial configuration with peak values reaching $T_b =0.5$.
In Fig.~\ref{fig:bulgetrix}, we show the evolution of the shape of the classical bulge. 
Initially the bulge is oblate ($T_b \sim 0$, see Fig.~\ref{fig:bulgetrix}); thereafter it evolves as a 
result of the angular momentum gain and change in the orbital structure. 
During the period of $0.17 - 0.56$~Gyr (roughly the dynamical phase) a considerable fraction of angular 
momentum is gained at resonances $\eta = -1$ corresponding to the Lagrange point orbits and $\eta =0.5$ (ILR) 
(see~Fig.~\ref{fig:resonances}). This angular momentum gain cause the bulge particles to move outwards 
and thereby producing a disky structure (see Fig.~\ref{fig:Bdenmaps}) in the outer parts of the bulge. 
In this period, essentially only $b/a$ changes while $c/a$ remains nearly constant.

Beyond $\sim 0.56$ Gyr, the ClBb forms in the bulge causing substantial changes in the bulge structure. As mentioned 
in section~\ref{sec:bulge-bar}, the ClBb heats \citep{Sahaetal2010} the bulge stars mainly in the central region 
and makes it thicker. We think that this heating due to ClBb is primarily responsible for subsequent changes in the $c/a$.
During the period from $1.5 - 2.1$~ Gyr, $b/a$ changes more than $c/a$. 
At $T=2.1$~Gyr, the small classical bulge is triaxial with $T_b= 0.48$. This suggests that more generally, 
fast rotating and triaxial bulges could have developed through the interaction of a strong bar and a small classical 
bulge in galaxies with low $B/D$ ratio. A more comprehensive analysis focusing on the role of the most important 
parameters such as the bulge mass and size will be presented in a future paper.

As shown in Section~\ref{sec:galaxymodels}, the initial velocity distribution in the classical bulge in our simulation 
is isotropic represented by a King model. As a result of the angular momentum influx and the readjustment of 
the orbits, the velocity structure changes during the evolution. We measure the deviation from isotropy in the velocity 
distribution by the anisotropy parameters defined as $\beta_{rz} = 1 - (\sigma_z/\sigma_r)^2$ and 
$\beta_{r\varphi} = 1 - (\sigma_{\varphi}/\sigma_r)^2$, where $\sigma_r$, $\sigma_{\phi}$ and $\sigma_z$ are the velocity dispersions
in the radial, azimuthal and vertical direction. Then $\beta_{rz} > 0$ denotes radial anisotropy and $\beta_{r\varphi} < 0$ means 
tangential anisotropy. In Fig.~\ref{fig:bulgeaniso}, we show the profiles of the radial anisotropy at four different epochs calculated from
all the particles in the classical bulge. We see that the classical bulge already becomes radially anisotropic at $T = 0.56$~ Gyr.
To understand the source of radial anisotropy, we have studied orbits in the classical bulge as clarified in Fig.~\ref{fig:resonances}. 
Fig.~\ref{fig:orbitaniso} shows radial variation of $\sigma_z^2/\sigma_r^2$ and $\sigma_{\phi}^2/\sigma_r^2$ computed from bulge particles
that are trapped at ILR ($\eta =0.5$) and from a group of resonant and non-resonant particles in the frequency range 
$-2.0 \le \eta \le-0.1$ separately. The particles that are at ILR resonance (2:1) follow x1-type orbits and they produce 
the radial anisotropy in the inner region of the classical bulge which host the ClBb. Whereas in the outer region, definite 
contributions to the radial anisotropy comes from the bulge particles in the frequency range $-2.0 \le \eta \le-0.1$. The particles that
are at resonance with the bar e.g., at $\eta = -2, -1$ follow regular orbits while the non-resonant particles are in stochastic orbits.  

\begin{figure}
\rotatebox{-90}{\includegraphics[height=8.5 cm]{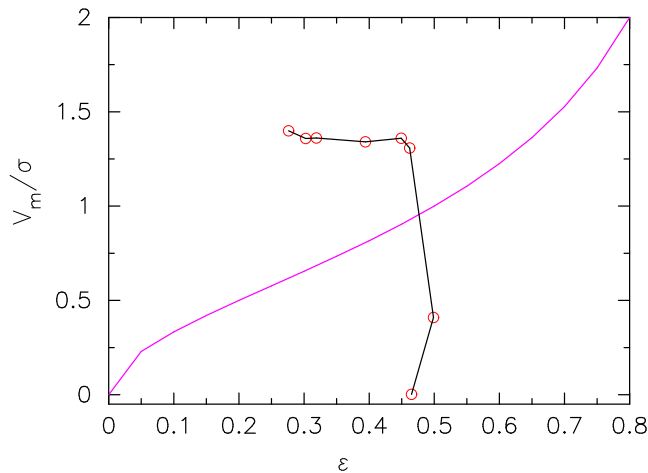}}
\caption{$V_m/\sigma - \epsilon$ relation for the small classical bulge alone. 
Initially the bulge is non-rotating but then it acquires angular momentum 
emitted by the bar and evolves into a fast rotating triaxial bulge. Each open 
circle represents an epoch in the simulation beginning at $T=~0$ (the bottom most point) 
. Subsequent circles are drwan at $0.25, 0.393, 0.56, 1.1, 1.52, 1.77, 2.1$ Gyr. 
The pink solid line is the reference isotropic rotator model.}
\label{fig:vsigeps}
\end{figure}

\subsection{$V_m/\sigma - \epsilon$ relation}
\label{sec:v-sig-epsa}
To quantify the degree of ordered motion in bulges and ellipticals and illuminate the difference 
between the two types of stellar systems, the $V_m/\sigma - \epsilon$ diagram 
relating  the ratio of rotational to random motions and the observed ellipticity 
($\epsilon$) was introduced \citep{Illingworth77}. It was shown that
bulges are, in general, fast rotators compared to bright elliptical galaxies \citep{KI1982, 
 Daviesetal1983, Cappellarietal2007, Morellietal2008}. Here, we focus
on the relation between the shape and the kinematics of the simulated low mass classical bulge that 
has been subject to the secular evolution driven by a strong bar. We show explicitly the evolutionary
track of this particular classical bulge in the $V_m/\sigma - \epsilon$ diagram below (Fig.~\ref{fig:vsigeps}). 

In observations, it is rather difficult to have an accurate measurement of the bulge rotation
velocity due to possible disk contamination. 
On the other hand, in simulations, it is rather straightforward to compute the velocity profile for the classical bulge alone
because it is possible to filter out the disk and halo components of the model galaxy. We determine, 
$V_m$ as the maximum of the azimuthally averaged rotational velocity of the bulge particles 
measured in the equatorial plane of the bulge; $\sigma$ is the mean velocity dispersion in the central region 
(calculated at $r \sim 0.5\times R_{1/2}$) of the bulge. 

For the bulge ellipticity, we use $c/a$ for edge-on view. Measuring the ellipticity for a bulge is a 
bit tricky because there can be strong radial variation in the ellipticity profile $\epsilon (r)$. 
Triaxiality could add another degree of complexity to such a measurement. Below we describe how the
bulge ellipticities are measured. 

In order to determine the intrinsic ellipticity, we first compute the moment-of-inertia tensor of the three 
dimensional mass distribution of the classical bulge and diagonalize it to 
obtain the principal moments and three orthogonal eigenvectors. The principal 
moments of inertia determine the intrinsic axis ratios of the inertia ellipsoid and the eigenvectors determine 
the orientation of the ellipsoid with respect to the co-ordinate space. Using the eigenvalues and eigenvectors, 
we determine the two axis ratios namely $b/a$ and $c/a$, where $a > b > c$ are the three semi-axes of the inertia 
ellipsoid. We have done this at the bulge half-mass radius containing $50$\% of the total bulge particles and at 
the radius containing about $90$\% of the total bulge particles. The ellipticity measurements for the classical 
bulge are nearly the same in both cases. In the following, we use the ellipticity at radii enclosing 
$\sim 90 \%$ of the total bulge particles and use the definition of the ellipticity of the bulge as 
$\epsilon = 1 - c/a$ when viewed edge-on. We have also performed isophote analysis using the 
ellipse-fitting routine from IRAF on a set of suitably rotated and inclined edge-on images 
(Fig.~\ref{fig:Bdenmaps}) of the classical bulge at different epochs during the evolution. We find a 
good agreement between the two different types of measurements of the bulge ellipticity. 

In Fig.~\ref{fig:vsigeps}, we show the $V_m/\sigma$ and $\epsilon$ values for the small classical bulge during the secular 
evolution. Each point in this diagram corresponds to a particular epoch
 during its evolution and when connected together they form its evolutionary track. This shows that the small 
classical bulge rotates significantly faster in the latter stages of evolution compared to the oblate isotropic rotator model, 
which can be approximated by \citep{Binney1978, K1982}:

\begin{equation}
V_m/\sigma \cong \sqrt{\frac{\epsilon}{1 - \epsilon}}.
\label{eq:vsigeps}
\end{equation}

\noindent The interpretation of Fig.~\ref{fig:vsigeps} is complicated by the fact that the classical bulge
is not an isolated stellar system, but interacts dynamically with the bar within a disk galaxy. The entire 
period of evolution of the classical bulge
can be broadly divided into two parts: one before the formation of the ClBb ($\sim 0.56$ Gyr) and the second after 
its formation. Before the formation of the ClBb, the bulge stars in the outer region gain a significant fraction 
of angular momentum emitted by the bar and move mainly outwards in radius. As a result 
of this, the values of $V_m/\sigma$ increase till $\sim 0.56$ Gyr while the axis ratio $c/a$ remains unchanged.   
In the second half of the evolution, the bulge stars are heated due to the ClBb by a factor of $\sim 1.5$ in velocity dispersion
 and this makes the inner region rounder. So $c/a$ ratio increases, making the ellipticity decrease 
considerably. The near saturation in the $V_m/\sigma$ towards the end of the simulation 
is connected with the fact that the rate of angular momentum gain by the bulge nearly saturates at these epochs 
(see Fig.~\ref{fig:spLz}, Fig.~\ref{fig:resonances}).   
A more detailed analysis on how the spinning up of the classical bulge 
depends on the various parameters of the bulge (bulge-to-disk mass ratio, size of the bulge, its central 
velocity dispersion) and disk (Toomre Q, bar strength, bar size) will be presented in a future paper.

\section{Discussion}
\label{sec:discussion}

The primary goal of this paper has been to describe a generic mechanism, the transfer of angular momentum
from a bar to an embedded classical bulge. We have shown that this mechanism is important for 
understanding the rotational motion of low mass classical bulges in the central regions of barred galaxies.
The growth rate and the strength of the bar are important factors for the mechanism to work efficiently.
One very interesting outcome of this process is the cylindrical rotation in the small classical bulge in 
the model studied here. Some possible observational implications and other issues are addressed below.

\subsection{Structural impact on the bulge}
During the secular evolution, a flat bar transforms into a boxy bulge. While the bar grows, buckles and evolves, a 
fraction of the angular momentum emitted by the bar is absorbed mainly in the outer parts of the embedded 
classical bulge in the galaxy. As a result, streaming motions are induced in the classical bulge (Fig.~\ref{fig:vcbulge}),
and the orbital structure changes, causing velocity anisotropies. 
After about half a Gyr, the ClBb forms in the inner regions of the bulge. The ClBb introduces a pattern rotation 
in the classical bulge which transforms into a rotating triaxial object. A comparision of 
Fig.~\ref{fig:barAMP} and Fig.~\ref{fig:bulgebarlm2} shows that beyond $\sim 0.8$ Gyr the ClBb essentially corotates with the 
boxy bulge formed out of the disk bar. Towards the end of the evolution, the model galaxy has a composite bulge: 
a superposition of the boxy bulge and a rotating triaxial classical bulge. 
At late times, the inner regions of the classical 
bulge become rounder as it is evident from the surface density 
maps (Fig.~\ref{fig:Bdenmaps}) for the classical bulge particles alone. We note that the stars in the classical bulge are 
 heated by a factor of $\sim 1.5$ in velocity dispersion within $2$ Gyrs. The ClBb maybe responsible 
for heating the bulge stars in a similar fashion as the bar heats the disk stars \citep{Sahaetal2010}.

\subsection{Boxy bulge and net cylindrical rotation}
It is widely accepted that cylindrical rotation is a characteristic feature of the kinematics of a boxy bulge
formed out of disk material, via the vertical buckling instability of the bar. Thus the presence of cylindrical 
rotation in the central regions of a galaxy may lead one to infer the presence of a boxy bulge that originated from 
the disk, without any need for a classical bulge in this galaxy. 

The work presented in this paper adds a new aspect to this simple picture. The cylindrical rotation could also include the 
stars of the classical bulge whose rotational properties have been modified by the interaction with the bar. One would 
measure the net cylindrical rotation of the stars in the combined boxy bulge and classical bulge. 
In absence of strong photometric evidence, other information such as from stellar populations and metallicity gradients 
would be needed to determine the presence of a small classical bulge. 
 
Although the pure kinematic modelling of the BRAVA data \citep{Shenetal2010} suggests only an 
upper limit on the mass of a classical bulge in the Milky Way, the measurements of the metallicity gradient
above the galactic plane \citep{zoccalietal2008} may indicate the presence of a classical bulge.
The upper limit on the total mass of the bulge (boxy bulge +classical bulge) in our model, including the remaining 
disk component in the boxy bulge region, is $\sim 1.46 \times 10^{10} M_{\odot}$. 
Of this, $0.29 \times 10^{10} M_{\odot}$ is in the classical bulge, and $\sim 1.17 \times 10^{10} M_{\odot}$ 
in the boxy bulge and the central disk. 
Since the classical bulge extends further above the galactic plane than the boxy bulge, the metallicity composition of the 
composite system would change with height. We plan to investigate this further and use our model to look for signatures 
of the classical bulge from the metallicity distribution in order to compare with observations of the Milky Way.
 
\subsection{Observing secular evolution through a classical bulge}
Previous studies mainly based on {\it N}-body simulations have focused on the
bar-halo interaction \citep{Athanassoula2002, WeinbergKatz2007a, villavargasetal2009, DubinskiBerentzen2009} and shown 
that a significant amount of angular momentum emitted by the bar is absorbed
in the dark matter halo. The angular momentum gained by the halo changes the internal structure of the
dark matter halo. Some authors have utilized this mechanism of angular momentum exchange to resolve the
cusp-core issue \citep{El-Zantetal2001, WeinbergKatz2002, Sellwood2008} in galaxies while others have focused on 
the halo-bar \citep{Holley-Bockelmannetal2005, colinetal2006, Athanassoula2007}. However, direct observational evidence for the halo-bar,
and hence direct observational verification of the ongoing secular evolution and angular momentum transfer can not be obtained unless 
dark matter is detected. Unlike the dark matter, it is possible to observe galactic bulge stars in detail, 
both the kinematics and stellar population parameters. It is thus possible to verify observationally the bar-bulge 
interaction and the resulting dynamical properties of the ClBb. Observational evidence for the ClBb could be a direct confirmation of the
angular momentum transfer and secular evolution in the galaxy. 

\section{Conclusions}
\label{sec:concl}
The secular processes driven by the bar not only restructure the disk, but also the other components
in the galaxy. Since the classical bulge is less massive here compared with the surrounding dark matter halo, the angular 
momentum gained by the classical bulge has a more significant effect on its evolution. The present work has shown
in considerable detail that the unavoidable gravitational interaction between these two components can have profound
implications for the structure of a low mass classical bulge, as highlighted below.

1. We have established that the main mechanism of angular momentum transport operating between the bar and the classical bulge is through 
resonances. The bulge particles gain angular momentum emitted by the bar through the bar's ILR ($\eta=0.5$),
and other resonances with $\eta = -1, -2$ and also through non-resonant orbits with $\eta \in (-0.4, 0.)$ during the dynamical 
phase when bar growth is rapid. Approximately $3/4$ of the net angular momentum is gained by the classical bulge during the 
dynamical phase where stochastic orbits contribute $\sim 30\%$.

2. The angular momentum gained by the initially non-rotating classical bulge sets the bulge particles in rotational motion.
The radial gradient in the rotational motion in the classical bulge is lower than in the boxy bulge. 
As time progresses the rotational velocity increases and nearly saturates at about 1 Gyr. At around this time, the inner 
regions of the classical bulge develop cylindrical rotation while the outer parts are still in differential rotation.

3. As a result of the angular momentum transfer, some of the bulge orbits are trapped by the rotating bar potential, a ClBb 
forms which essentially cororates with the bar, and the classical bulge transforms into a triaxial, anisotropic object where
trapped orbits contributes to the radial anisotropy.   

4. Towards the end of the secular evolution, the model galaxy has a composite bulge which is a superposition of the boxy bulge formed 
out of the disk material and the rotating triaxial, low mass classical bulge. The stars in the composite bulge rotates cylindrically.  
From an observational perspective, one would need other tracers such as metallicity gradient, stellar population parameters along with the
kinematics to reliably determine the presence of such a low mass classical bulge embedded in the boxy bulge. 


\noindent{\bf Acknowledgements}

\noindent K.S. acknowledges support from the Alexander von Humboldt Foundation. It is a pleasure to thank 
Lodovico Coccato for his generous help with IRAF, Jerry Sellwood for letting us use his potential solver code, 
and Francoise Combes for valuable inputs. The authors thank the anonymous referee for useful comments.



\begin{thebibliography}{}
\bibitem[\protect\citeauthoryear{{Aguerri}, {Balcells} \& {Peletier}}{{Aguerri}
  et~al.}{2001}]{Aguerrietal2001}
{Aguerri} J.~A.~L.,  {Balcells} M.,    {Peletier} R.~F.,  2001, \aap, 367, 428

\bibitem[\protect\citeauthoryear{{Ann}}{{Ann}}{1995}]{Ann1995}
{Ann} H.~B.,  1995, Journal of Korean Astronomical Society, 28, 209

\bibitem[\protect\citeauthoryear{{Athanassoula}}{{Athanassoula}}{2002}]{Athana%
ssoula2002}
{Athanassoula} E.,  2002, \apjl, 569, L83

\bibitem[\protect\citeauthoryear{{Athanassoula}}{{Athanassoula}}{2003}]{Athana%
ssoula2003}
{Athanassoula} E.,  2003, \mnras, 341, 1179

\bibitem[\protect\citeauthoryear{{Athanassoula}}{{Athanassoula}}{2005}]{Athana%
ssoula2005}
{Athanassoula} E.,  2005, \mnras, 358, 1477

\bibitem[\protect\citeauthoryear{{Athanassoula}}{{Athanassoula}}{2007}]{Athana%
ssoula2007}
{Athanassoula} E.,  2007, \mnras, 377, 1569

\bibitem[\protect\citeauthoryear{{Athanassoula} \& {Misiriotis}}{{Athanassoula}
  \& {Misiriotis}}{2002}]{Athanamisi2002}
{Athanassoula} E.,  {Misiriotis} A.,  2002, \mnras, 330, 35

\bibitem[\protect\citeauthoryear{{Baugh}, {Cole} \& {Frenk}}{{Baugh}
  et~al.}{1996}]{Baughetal1996}
{Baugh} C.~M.,  {Cole} S.,    {Frenk} C.~S.,  1996, \mnras, 283, 1361

\bibitem[\protect\citeauthoryear{{Bertola}, {Vietri} \& {Zeilinger}}{{Bertola}
  et~al.}{1991}]{Bertolaetal1991}
{Bertola} F.,  {Vietri} M.,    {Zeilinger} W.~W.,  1991, \apjl, 374, L13

\bibitem[\protect\citeauthoryear{{Binney}}{{Binney}}{1978}]{Binney1978}
{Binney} J.,  1978, \mnras, 183, 501

\bibitem[\protect\citeauthoryear{{Binney} \& {Spergel}}{{Binney} \&
  {Spergel}}{1982}]{BinneySpergel1982}
{Binney} J.,  {Spergel} D.,  1982, \apj, 252, 308

\bibitem[\protect\citeauthoryear{{Binney} \& {Tremaine}}{{Binney} \&
  {Tremaine}}{1987}]{BT1987}
{Binney} J.,  {Tremaine} S.,  1987, {Galactic dynamics}

\bibitem[\protect\citeauthoryear{{Bournaud}, {Jog} \& {Combes}}{{Bournaud}
  et~al.}{2007}]{Bournaudetal2007}
{Bournaud} F.,  {Jog} C.~J.,    {Combes} F.,  2007, \aap, 476, 1179

\bibitem[\protect\citeauthoryear{{Cappellari}, {Emsellem}, {Bacon}, {Bureau},
  {Davies}, {de Zeeuw}, {Falc{\'o}n-Barroso}, {Krajnovi{\'c}}, {Kuntschner},
  {McDermid}, {Peletier}, {Sarzi}, {van den Bosch} \& {van de
  Ven}}{{Cappellari} et~al.}{2007}]{Cappellarietal2007}
{Cappellari} M.,  {Emsellem} E.,  {Bacon} R.,  {Bureau} M.,  {Davies} R.~L.,
  {de Zeeuw} P.~T.,  {Falc{\'o}n-Barroso} J.,  {Krajnovi{\'c}} D.,
  {Kuntschner} H.,  {McDermid} R.~M.,  {Peletier} R.~F.,  {Sarzi} M.,  {van den
  Bosch} R.~C.~E.,    {van de Ven} G.,  2007, \mnras, 379, 418

\bibitem[\protect\citeauthoryear{{Col{\'{\i}}n}, {Valenzuela} \&
  {Klypin}}{{Col{\'{\i}}n} et~al.}{2006}]{colinetal2006}
{Col{\'{\i}}n} P.,  {Valenzuela} O.,    {Klypin} A.,  2006, \apj, 644, 687

\bibitem[\protect\citeauthoryear{{Combes}}{{Combes}}{2009}]{Combes2009}
{Combes} F.,  2009, in {S.~Jogee, I.~Marinova, L.~Hao, \& G.~A.~Blanc} ed.,
  Galaxy Evolution: Emerging Insights and Future Challenges Vol.~419 of
  Astronomical Society of the Pacific Conference Series, {Secular Evolution and
  the Assembly of Bulges}.
pp 31--+

\bibitem[\protect\citeauthoryear{{Combes} \& {Sanders}}{{Combes} \&
  {Sanders}}{1981}]{CombesSanders1981}
{Combes} F.,  {Sanders} R.~H.,  1981, \aap, 96, 164

\bibitem[\protect\citeauthoryear{{Davies}, {Efstathiou}, {Fall}, {Illingworth}
  \& {Schechter}}{{Davies} et~al.}{1983}]{Daviesetal1983}
{Davies} R.~L.,  {Efstathiou} G.,  {Fall} S.~M.,  {Illingworth} G.,
  {Schechter} P.~L.,  1983, \apj, 266, 41

\bibitem[\protect\citeauthoryear{{Debattista} \& {Sellwood}}{{Debattista} \&
  {Sellwood}}{2000}]{DebattistaSellwood2000}
{Debattista} V.~P.,  {Sellwood} J.~A.,  2000, \apj, 543, 704

\bibitem[\protect\citeauthoryear{{Dubinski}, {Berentzen} \&
  {Shlosman}}{{Dubinski} et~al.}{2009}]{DubinskiBerentzen2009}
{Dubinski} J.,  {Berentzen} I.,    {Shlosman} I.,  2009, \apj, 697, 293

\bibitem[\protect\citeauthoryear{{Eggen}, {Lynden-Bell} \& {Sandage}}{{Eggen}
  et~al.}{1962}]{Eggenetal1962}
{Eggen} O.~J.,  {Lynden-Bell} D.,    {Sandage} A.~R.,  1962, \apj, 136, 748

\bibitem[\protect\citeauthoryear{{El-Zant}, {Shlosman} \& {Hoffman}}{{El-Zant}
  et~al.}{2001}]{El-Zantetal2001}
{El-Zant} A.,  {Shlosman} I.,    {Hoffman} Y.,  2001, \apj, 560, 636

\bibitem[\protect\citeauthoryear{{Elmegreen}, {Bournaud} \&
  {Elmegreen}}{{Elmegreen} et~al.}{2008}]{Elmegreenetal2008}
{Elmegreen} B.~G.,  {Bournaud} F.,    {Elmegreen} D.~M.,  2008, \apj, 688, 67

\bibitem[\protect\citeauthoryear{{Erwin}}{{Erwin}}{2008}]{Erwin2008}
{Erwin} P.,  2008, in {M.~Bureau, E.~Athanassoula, \& B.~Barbuy} ed., IAU
  Symposium Vol.~245 of IAU Symposium, {The coexistence of classical bulges and
  disky pseudobulges in early-type disk galaxies}.
pp 113--116

\bibitem[\protect\citeauthoryear{{Evans}}{{Evans}}{1993}]{Evans1993}
{Evans} N.~W.,  1993, \mnras, 260, 191

\bibitem[\protect\citeauthoryear{{Fall} \& {Efstathiou}}{{Fall} \&
  {Efstathiou}}{1980}]{FallEfstathiou1980}
{Fall} S.~M.,  {Efstathiou} G.,  1980, \mnras, 193, 189

\bibitem[\protect\citeauthoryear{{Franx}, {Illingworth} \& {de Zeeuw}}{{Franx}
  et~al.}{1991}]{Franxetal1991}
{Franx} M.,  {Illingworth} G.,    {de Zeeuw} T.,  1991, \apj, 383, 112

\bibitem[\protect\citeauthoryear{{Gadotti}}{{Gadotti}}{2009}]{Gadotti2009}
{Gadotti} D.~A.,  2009, \mnras, 393, 1531

\bibitem[\protect\citeauthoryear{{Gerhard}, {Vietri} \& {Kent}}{{Gerhard}
  et~al.}{1989}]{Gerhardetal1989}
{Gerhard} O.~E.,  {Vietri} M.,    {Kent} S.~M.,  1989, \apjl, 345, L33

\bibitem[\protect\citeauthoryear{{Hernquist} \& {Weinberg}}{{Hernquist} \&
  {Weinberg}}{1992}]{HernquistWeinberg1992}
{Hernquist} L.,  {Weinberg} M.~D.,  1992, \apj, 400, 80

\bibitem[\protect\citeauthoryear{{Holley-Bockelmann}, {Weinberg} \&
  {Katz}}{{Holley-Bockelmann} et~al.}{2005}]{Holley-Bockelmannetal2005}
{Holley-Bockelmann} K.,  {Weinberg} M.,    {Katz} N.,  2005, \mnras, 363, 991

\bibitem[\protect\citeauthoryear{{Hopkins}, {Bundy}, {Croton}, {Hernquist},
  {Keres}, {Khochfar}, {Stewart}, {Wetzel} \& {Younger}}{{Hopkins}
  et~al.}{2010}]{Hopkinsetal2010}
{Hopkins} P.~F.,  {Bundy} K.,  {Croton} D.,  {Hernquist} L.,  {Keres} D.,
  {Khochfar} S.,  {Stewart} K.,  {Wetzel} A.,    {Younger} J.~D.,  2010, \apj,
  715, 202

\bibitem[\protect\citeauthoryear{{Hopkins}, {Cox}, {Younger} \&
  {Hernquist}}{{Hopkins} et~al.}{2009}]{Hopkinsetal2009}
{Hopkins} P.~F.,  {Cox} T.~J.,  {Younger} J.~D.,    {Hernquist} L.,  2009,
  \apj, 691, 1168

\bibitem[\protect\citeauthoryear{{Illingworth}}{{Illingworth}}{1977}]{Illingwo%
rth77}
{Illingworth} G.,  1977, \apjl, 218, L43

\bibitem[\protect\citeauthoryear{{Immeli}, {Samland}, {Gerhard} \&
  {Westera}}{{Immeli} et~al.}{2004}]{Immelietal2004}
{Immeli} A.,  {Samland} M.,  {Gerhard} O.,    {Westera} P.,  2004, \aap, 413,
  547

\bibitem[\protect\citeauthoryear{{Jesseit}, {Naab} \& {Burkert}}{{Jesseit}
  et~al.}{2005}]{Jesseitetal2005}
{Jesseit} R.,  {Naab} T.,    {Burkert} A.,  2005, \mnras, 360, 1185

\bibitem[\protect\citeauthoryear{{Katz}, {Keres}, {Dave} \& {Weinberg}}{{Katz}
  et~al.}{2003}]{Katzetal2003}
{Katz} N.,  {Keres} D.,  {Dave} R.,    {Weinberg} D.~H.,  2003, in
  {J.~L.~Rosenberg \& M.~E.~Putman} ed., The IGM/Galaxy Connection. The
  Distribution of Baryons at z=0 Vol.~281 of Astrophysics and Space Science
  Library, {How Do Galaxies Get Their Gas?}.
pp 185--+

\bibitem[\protect\citeauthoryear{{Kauffmann}, {White} \&
  {Guiderdoni}}{{Kauffmann} et~al.}{1993}]{Kauffmanetal1993}
{Kauffmann} G.,  {White} S.~D.~M.,    {Guiderdoni} B.,  1993, \mnras, 264, 201

\bibitem[\protect\citeauthoryear{{Kere{\v s}}, {Katz}, {Fardal}, {Dav{\'e}} \&
  {Weinberg}}{{Kere{\v s}} et~al.}{2009}]{Keresetal2009}
{Kere{\v s}} D.,  {Katz} N.,  {Fardal} M.,  {Dav{\'e}} R.,    {Weinberg} D.~H.,
   2009, \mnras, 395, 160

\bibitem[\protect\citeauthoryear{{King}}{{King}}{1966}]{King1966}
{King} I.~R.,  1966, \aj, 71, 64

\bibitem[\protect\citeauthoryear{{Kormendy}}{{Kormendy}}{1982}]{K1982}
{Kormendy} J.,  1982, \apj, 257, 75

\bibitem[\protect\citeauthoryear{{Kormendy} \& {Illingworth}}{{Kormendy} \&
  {Illingworth}}{1982}]{KI1982}
{Kormendy} J.,  {Illingworth} G.,  1982, \apj, 256, 460

\bibitem[\protect\citeauthoryear{{Kormendy} \& {Kennicutt} Jr.}{{Kormendy} \&
  {Kennicutt}}{2004}]{KormendyKennicut2004}
{Kormendy} J.,  {Kennicutt} Jr. R.~C.,  2004, \araa, 42, 603

\bibitem[\protect\citeauthoryear{{Kuijken} \& {Dubinski}}{{Kuijken} \&
  {Dubinski}}{1995}]{KD1995}
{Kuijken} K.,  {Dubinski} J.,  1995, \mnras, 277, 1341

\bibitem[\protect\citeauthoryear{{Laurikainen}, {Salo}, {Buta} \&
  {Vasylyev}}{{Laurikainen} et~al.}{2004}]{Laurikainenetal2004}
{Laurikainen} E.,  {Salo} H.,  {Buta} R.,    {Vasylyev} S.,  2004, \mnras, 355,
  1251

\bibitem[\protect\citeauthoryear{{Lynden-Bell} \& {Kalnajs}}{{Lynden-Bell} \&
  {Kalnajs}}{1972}]{Lynden-BellKalnajs1972}
{Lynden-Bell} D.,  {Kalnajs} A.~J.,  1972, \mnras, 157, 1

\bibitem[\protect\citeauthoryear{{Marinova} \& {Jogee}}{{Marinova} \&
  {Jogee}}{2007}]{MarinovaJogee2007}
{Marinova} I.,  {Jogee} S.,  2007, \apj, 659, 1176

\bibitem[\protect\citeauthoryear{{Martinez-Valpuesta} \&
  {Shlosman}}{{Martinez-Valpuesta} \& {Shlosman}}{2004}]{MV2004}
{Martinez-Valpuesta} I.,  {Shlosman} I.,  2004, \apjl, 613, L29

\bibitem[\protect\citeauthoryear{{Martinez-Valpuesta}, {Shlosman} \&
  {Heller}}{{Martinez-Valpuesta} et~al.}{2006}]{MV2006}
{Martinez-Valpuesta} I.,  {Shlosman} I.,    {Heller} C.,  2006, \apj, 637, 214

\bibitem[\protect\citeauthoryear{{McMillan} \& {Dehnen}}{{McMillan} \&
  {Dehnen}}{2007}]{McMillan2007}
{McMillan} P.~J.,  {Dehnen} W.,  2007, \mnras, 378, 541

\bibitem[\protect\citeauthoryear{{M{\'e}ndez-Abreu}, {Simonneau}, {Aguerri} \&
  {Corsini}}{{M{\'e}ndez-Abreu} et~al.}{2010}]{Mendezetal2010}
{M{\'e}ndez-Abreu} J.,  {Simonneau} E.,  {Aguerri} J.~A.~L.,    {Corsini}
  E.~M.,  2010, \aap, 521, A71+

\bibitem[\protect\citeauthoryear{{Men{\'e}ndez-Delmestre}, {Sheth},
  {Schinnerer}, {Jarrett} \& {Scoville}}{{Men{\'e}ndez-Delmestre}
  et~al.}{2007}]{MenendezDelmestreetal2007}
{Men{\'e}ndez-Delmestre} K.,  {Sheth} K.,  {Schinnerer} E.,  {Jarrett} T.~H.,
   {Scoville} N.~Z.,  2007, \apj, 657, 790

\bibitem[\protect\citeauthoryear{{Mo}, {Mao} \& {White}}{{Mo}
  et~al.}{1998}]{MoMaowhite1998}
{Mo} H.~J.,  {Mao} S.,    {White} S.~D.~M.,  1998, \mnras, 295, 319

\bibitem[\protect\citeauthoryear{{Morelli}, {Pompei}, {Pizzella},
  {M{\'e}ndez-Abreu}, {Corsini}, {Coccato}, {Saglia}, {Sarzi} \&
  {Bertola}}{{Morelli} et~al.}{2008}]{Morellietal2008}
{Morelli} L.,  {Pompei} E.,  {Pizzella} A.,  {M{\'e}ndez-Abreu} J.,  {Corsini}
  E.~M.,  {Coccato} L.,  {Saglia} R.~P.,  {Sarzi} M.,    {Bertola} F.,  2008,
  \mnras, 389, 341

\bibitem[\protect\citeauthoryear{{Nieto} \& {Bender}}{{Nieto} \&
  {Bender}}{1989}]{NietoBender89}
{Nieto} J.-L.,  {Bender} R.,  1989, \aap, 215, 266

\bibitem[\protect\citeauthoryear{{Nowak}, {Thomas}, {Erwin}, {Saglia}, {Bender}
  \& {Davies}}{{Nowak} et~al.}{2010}]{Nowaketal2010}
{Nowak} N.,  {Thomas} J.,  {Erwin} P.,  {Saglia} R.~P.,  {Bender} R.,
  {Davies} R.~I.,  2010, \mnras, 403, 646

\bibitem[\protect\citeauthoryear{{Pfenniger} \& {Norman}}{{Pfenniger} \&
  {Norman}}{1990}]{PfennigerNorman1990}
{Pfenniger} D.,  {Norman} C.,  1990, \apj, 363, 391

\bibitem[\protect\citeauthoryear{{Raha}, {Sellwood}, {James} \& {Kahn}}{{Raha}
  et~al.}{1991}]{Rahaetal1991}
{Raha} N.,  {Sellwood} J.~A.,  {James} R.~A.,    {Kahn} F.~D.,  1991, \nat,
  352, 411

\bibitem[\protect\citeauthoryear{{Saha}, {Tseng} \& {Taam}}{{Saha}
  et~al.}{2010}]{Sahaetal2010}
{Saha} K.,  {Tseng} Y.,    {Taam} R.~E.,  2010, \apj, 721, 1878

\bibitem[\protect\citeauthoryear{{Sellwood}}{{Sellwood}}{1981}]{Sellwood1981}
{Sellwood} J.~A.,  1981, \aap, 99, 362

\bibitem[\protect\citeauthoryear{{Sellwood}}{{Sellwood}}{2008}]{Sellwood2008}
{Sellwood} J.~A.,  2008, \apj, 679, 379

\bibitem[\protect\citeauthoryear{{Sellwood} \& {Debattista}}{{Sellwood} \&
  {Debattista}}{2006}]{SellwoodDebattista2006}
{Sellwood} J.~A.,  {Debattista} V.~P.,  2006, \apj, 639, 868

\bibitem[\protect\citeauthoryear{{Sellwood} \& {Evans}}{{Sellwood} \&
  {Evans}}{2001}]{SellwoodEvans2001}
{Sellwood} J.~A.,  {Evans} N.~W.,  2001, \apj, 546, 176

\bibitem[\protect\citeauthoryear{{Sellwood} \& {Valluri}}{{Sellwood} \&
  {Valluri}}{1997}]{SellwoodValluri1997}
{Sellwood} J.~A.,  {Valluri} M.,  1997, \mnras, 287, 124

\bibitem[\protect\citeauthoryear{{Sellwood} \& {Wilkinson}}{{Sellwood} \&
  {Wilkinson}}{1993}]{SellwoodWilkinson1993}
{Sellwood} J.~A.,  {Wilkinson} A.,  1993, Reports on Progress in Physics, 56,
  173

\bibitem[\protect\citeauthoryear{{Shen}, {Rich}, {Kormendy}, {Howard}, {De
  Propris} \& {Kunder}}{{Shen} et~al.}{2010}]{Shenetal2010}
{Shen} J.,  {Rich} R.~M.,  {Kormendy} J.,  {Howard} C.~D.,  {De Propris} R.,
  {Kunder} A.,  2010, \apjl, 720, L72

\bibitem[\protect\citeauthoryear{{Springel} \& {Hernquist}}{{Springel} \&
  {Hernquist}}{2005}]{SpringelHernquist2005}
{Springel} V.,  {Hernquist} L.,  2005, \apjl, 622, L9

\bibitem[\protect\citeauthoryear{{Springel}, {Yoshida} \& {White}}{{Springel}
  et~al.}{2001}]{Springeletal2001}
{Springel} V.,  {Yoshida} N., {White} S.~D.~M.,  2001, \na, 6, 79

\bibitem[\protect\citeauthoryear{{Stark}}{{Stark}}{1977}]{Stark77}
{Stark} A.~A.,  1977, \apj, 213, 368

\bibitem[\protect\citeauthoryear{{Toomre}}{{Toomre}}{1981}]{Toomre1981}
{Toomre} A.,  1981, in {S.~M.~Fall \& D.~Lynden-Bell} ed., Structure and
  Evolution of Normal Galaxies {What amplifies the spirals}.
pp 111--136

\bibitem[\protect\citeauthoryear{{Tremaine} \& {Weinberg}}{{Tremaine} \&
  {Weinberg}}{1984}]{TremaineWeinberg1984}
{Tremaine} S.,  {Weinberg} M.~D.,  1984, \mnras, 209, 729

\bibitem[\protect\citeauthoryear{{Villa-Vargas}, {Shlosman} \&
  {Heller}}{{Villa-Vargas} et~al.}{2009}]{villavargasetal2009}
{Villa-Vargas} J.,  {Shlosman} I.,    {Heller} C.,  2009, \apj, 707, 218

\bibitem[\protect\citeauthoryear{{Weinberg}}{{Weinberg}}{1985}]{Weinberg1985}
{Weinberg} M.~D.,  1985, \mnras, 213, 451

\bibitem[\protect\citeauthoryear{{Weinberg} \& {Katz}}{{Weinberg} \&
  {Katz}}{2002}]{WeinbergKatz2002}
{Weinberg} M.~D.,  {Katz} N.,  2002, \apj, 580, 627

\bibitem[\protect\citeauthoryear{{Weinberg} \& {Katz}}{{Weinberg} \&
  {Katz}}{2007a}]{WeinbergKatz2007a}
{Weinberg} M.~D.,  {Katz} N.,  2007a, \mnras, 375, 425

\bibitem[\protect\citeauthoryear{{Weinberg} \& {Katz}}{{Weinberg} \&
  {Katz}}{2007b}]{WeinbergKatz2007b}
{Weinberg} M.~D.,  {Katz} N.,  2007b, \mnras, 375, 460

\bibitem[\protect\citeauthoryear{{White} \& {Rees}}{{White} \&
  {Rees}}{1978}]{WhiteRees1978}
{White} S.~D.~M.,  {Rees} M.~J.,  1978, \mnras, 183, 341

\bibitem[\protect\citeauthoryear{{Widrow}, {Pym} \& {Dubinski}}{{Widrow}
  et~al.}{2008}]{Widrowetal2008}
{Widrow} L.~M.,  {Pym} B.,    {Dubinski} J.,  2008, \apj, 679, 1239

\bibitem[\protect\citeauthoryear{{Zoccali}}{{Zoccali}}{2010}]{zoccali2010}
{Zoccali} M.,  2010, in {K.~Cunha, M.~Spite, \& B.~Barbuy} ed., IAU Symposium
  Vol.~265 of IAU Symposium, {The Stellar Population of the Galactic Bulge}.
pp 271--278

\bibitem[\protect\citeauthoryear{{Zoccali}, {Hill}, {Lecureur}, {Barbuy},
  {Renzini}, {Minniti}, {G{\'o}mez} \& {Ortolani}}{{Zoccali}
  et~al.}{2008}]{zoccalietal2008}
{Zoccali} M.,  {Hill} V.,  {Lecureur} A.,  {Barbuy} B.,  {Renzini} A.,
  {Minniti} D.,  {G{\'o}mez} A.,    {Ortolani} S.,  2008, \aap, 486, 177

\end{thebibliography}
\end{document}